\documentclass{aastex63}
\usepackage{comment}
\usepackage{lineno}
\usepackage{array,multirow}

\newcommand{\TextUnderscore}{\rule{.5em}{.3pt}}

\shorttitle{Radial velocity dispersion profile of NGC 6440}
\shortauthors{Leanza et al.}

\graphicspath{{./}{figures/}}

\begin{document}

\title{The ESO-VLT MIKiS survey reloaded: exploring the internal kinematics of 
NGC 6440\footnote{Based on observations
  collected at the European Southern Observatory, Cerro Paranal
  (Chile), under proposal 193.D-0232 (PI: Ferraro),
  195.D-0750 (PI: Ferraro), 60.A-9489 (PI: Ferraro) and 093.D-0319 (PI: Lanzoni) }}

\correspondingauthor{Silvia Leanza}
\email{silvia.leanza2@unibo.it}

\author[0000-0001-9545-5291]{Silvia Leanza}
\affil{Dipartimento di Fisica e Astronomia, Universit\`a di Bologna, Via Gobetti 93/2 I-40129 Bologna, Italy}
\affil{INAF-Osservatorio di Astrofisica e Scienze dello Spazio di Bologna, Via Gobetti 93/3 I-40129 Bologna, Italy}

\author[0000-0002-7104-2107]{Cristina Pallanca}
\affil{Dipartimento di Fisica e Astronomia, Universit\`a di Bologna, Via Gobetti 93/2 I-40129 Bologna, Italy}
\affil{INAF-Osservatorio di Astrofisica e Scienze dello Spazio di Bologna, Via Gobetti 93/3 I-40129 Bologna, Italy}

\author[0000-0002-2165-8528]{Francesco R. Ferraro}
\affil{Dipartimento di Fisica e Astronomia, Universit\`a di Bologna, Via Gobetti 93/2 I-40129 Bologna, Italy}
\affil{INAF-Osservatorio di Astrofisica e Scienze dello Spazio di Bologna, Via Gobetti 93/3 I-40129 Bologna, Italy}

\author[0000-0001-5613-4938]{Barbara Lanzoni}
\affil{Dipartimento di Fisica e Astronomia, Universit\`a di Bologna, Via Gobetti 93/2 I-40129 Bologna, Italy}
\affil{INAF-Osservatorio di Astrofisica e Scienze dello Spazio di Bologna, Via Gobetti 93/3 I-40129 Bologna, Italy}

\author[0000-0003-4237-4601]{Emanuele Dalessandro}
\affil{INAF-Osservatorio di Astrofisica e Scienze dello Spazio di Bologna, Via Gobetti 93/3 I-40129 Bologna, Italy}

\author[0000-0002-5038-3914]{Mario Cadelano}
\affil{Dipartimento di Fisica e Astronomia, Universit\`a di Bologna, Via Gobetti 93/2 I-40129 Bologna, Italy}
\affil{INAF-Osservatorio di Astrofisica e Scienze dello Spazio di Bologna, Via Gobetti 93/3 I-40129 Bologna, Italy}

\author[0000-0003-2742-6872]{Enrico Vesperini}
\affil{Department of Astronomy, Indiana University, Bloomington, IN, 47401, USA}

\author[0000-0002-6040-5849]{Livia Origlia}
\affil{INAF-Osservatorio di Astrofisica e Scienze dello Spazio di Bologna, Via Gobetti 93/3 I-40129 Bologna, Italy}

\author[0000-0001-9158-8580]{Alessio Mucciarelli}
\affil{Dipartimento di Fisica e Astronomia, Universit\`a di Bologna, Via Gobetti 93/2 I-40129 Bologna, Italy}
\affil{INAF-Osservatorio di Astrofisica e Scienze dello Spazio di Bologna, Via Gobetti 93/3 I-40129 Bologna, Italy}

\author[0000-0002-6092-7145]{Elena Valenti}
\affil{European Southern Observatory, Karl-Schwarzschild-Strasse 2, 85748 Garching bei Munchen, Germany}
\affil{Excellence Cluster ORIGINS, Boltzmann-Strasse 2, D-85748 Garching Bei Munchen, Germany}
 




\begin{abstract}
In the context of the ESO-VLT Multi-Instrument Kinematic Survey
(MIKiS) of Galactic globular clusters, here we present the
line-of-sight velocity dispersion profile of NGC 6440, a massive
globular cluster located in the Galactic bulge. By combining the data
acquired with four different spectrographs, we obtained
the radial velocity of a sample of $\sim 1800$
individual stars distributed over the entire cluster extension,
from $\sim 0.1\arcsec$ to $778\arcsec$ from the center. 
Using a properly selected sample of member stars with the most 
reliable radial velocity measures, we derived the velocity dispersion profile
up to $250\arcsec$ from the center. The profile
is well described by the same King model that best
fits the projected star density distribution, with a constant 
inner plateau (at $\sigma_0\sim 12$ km s$^{-1}$) and no evidence of a
central cusp or other significant deviations.  
Our data allowed to study the presence of rotation only in the
innermost regions of the cluster ($r<5\arcsec$), revealing 
a well-defined pattern of ordered rotation with a position angle of the
rotation axis of $\sim 132 \pm 2 \arcdeg$ and an amplitude of $\sim3$ km
s$^{-1}$ (corresponding to $V_{rot} / \sigma_0 \sim 0.3$).
Also, a ﬂattening of the system qualitatively consistent 
with the rotation signal has been detected in the central region.
\end{abstract}



\section{Introduction} \label{sec:intro}
The ESO-VLT Multi-Instrument Kinematic Survey (hereafter the MIKiS
survey; \citealt{ferraro+18a, ferraro+18b}) has been specifically
designed to characterize the kinematical properties of a
representative sample of Galactic globular clusters (GGCs) in
different dynamical evolutionary stages.  The approach proposed in
MIKiS is to derive both the velocity dispersion and the rotation
profiles from the line-of-sight velocities of a statistically
significant sample of individual stars distributed over the entire
radial extension of each investigated stellar system. To this end, the
spectroscopic capabilities of different instruments located
at the ESO Very Large Telescope (VLT) are used: the survey
was designed to take advantage of the adaptive-optics (AO)
assisted integral-field spectrograph SINFONI, the multi-object
integral-field spectrograph KMOS, and the multi-object fiber-fed
spectrograph FLAMES/GIRAFFE, and it has been recently complemented
with a series of specific proposals and an ongoing Large Program
(106.21N5, PI: Ferraro) which exploits the remarkable performances of
the AO-assisted integral-field spectrograph MUSE.

In particular, the powerful combination of the two AO-assisted
integral-field spectrographs SINFONI and MUSE allows an unprecedented
exploration of the kinematics of the innermost GGC core regions,
reaching a spatial resolution comparable with that obtained from
Hubble Space Telescope (HST) observations.  Indeed the central
portions of collisional stellar systems like the GGCs are the most
intriguing regions where recurrent stellar interactions are expected
to generate exotic objects, like interacting binaries, blue
stragglers, millisecond pulsars \citep{bailyn95, pooley+03, ransom+05,
  ferraro+97, ferraro+01, ferraro+18c}.  Moreover, even the
long-sought intermediate-mass black holes \citep[IMBHs;
  e.g.][]{portegies+04,giersz+15} might populate the central regions 
of GCs according to the extrapolation of the
``Magorrian relation'' \citep{magorrian+98}.

The MIKiS survey is part of a long-term project (Cosmic-Lab) aimed at
performing a comprehensive study of a sample of representative
GGCs. Thus, the kinematical information provided by MIKiS is combined
with additional key properties of each system derived from
complementary observations as: (1) plane-of-the-sky kinematics
obtained from the stellar proper motions measured with HST and GAIA
\citep[see, e.g.,][]{raso+20,leanza+22}; (2) updated structural
parameters obtained from high-quality density profiles derived from
star counts, instead of surface brightness \citep[e.g.,][]{miocchi+13,
  lanzoni+07, lanzoni+10, lanzoni+19, pallanca+21}; (3) properties of
the populations of stellar exotica \citep{ferraro+01, ferraro+03,
  ferraro+15, ferraro+16, pallanca+10, pallanca+13, pallanca+14,
  pallanca+17, cadelano+17, cadelano+18, cadelano+20} and their
connection with the dynamical evolution of the parent cluster
\citep[see][]{ferraro+06, ferraro+09a, ferraro+12,ferraro+18a,
  ferraro+19, lanzoni+16}.  In principle, this approach can provide
the full characterization of the investigated stellar systems.

In this paper, we present the velocity dispersion profile and 
the detection of internal rotation for NGC 6440,
a massive, metal-rich ([Fe/H]$\sim -0.56$,
\citealt{origlia+97,origlia+08}), and highly extincted
\citep[$E(B-V)=1.15$,][see also \citealt{pallanca+19}]{valenti+04,valenti+07} GC, located in the
Milky Way bulge, 1.3 kpc away from the center of the Galaxy
\citep[][]{harris+96}.  This system has been subject to a detailed
analysis by our group because it was suspected \citep{mauro+12} to
have properties similar to those detected in Terzan 5 and Liller 1,
two massive clusters in the bulge direction that are suspected to be
the relics of the primordial assembling process of the Galactic bulge
and define a new class of stellar systems named ``Bulge Fossil
Fragments'' \citep[see][]{ferraro+09b, ferraro+16, ferraro+21}.  Those
studies allowed us to re-determine with an improved level of accuracy
the overall characteristics of NGC 6440, thus redesigning a sort of
new identity card of the cluster.  A high-resolution extinction map
has been obtained \citep[see][]{pallanca+19} and used to correct the
effects of the strong differential reddening in the direction of the
cluster. This allowed us to derive a high-precision, differential
reddening-corrected, proper motion-selected color-magnitude diagram
(CMD), from which a new determination of the cluster age has been
obtained \citep[see][]{pallanca+21}. This dataset has also been used
to derive an accurate star density profile from which new structural
parameters and characteristic relaxation times have been derived
\citep[see][]{pallanca+21}.  With regard to the study of the exotic
populations, we provided the identification of the optical companion
to an accreting millisecond X-ray pulsar \citep[see][]{cadelano+17}.

The present paper is organized as follows. In Section \ref{sec:obs} we
present the observations and describe the procedures performed 
for the data reduction.
In Section \ref{sec:analysis} we discuss the selection of the samples,
the methods to determine the radial velocities (RVs), and the strategy 
adopted to homogenize the different datasets available. The results
are presented in Section \ref{sec:results}, while Section
\ref{sec:discussion} is devoted to the discussion and conclusions.

\section{Observations and data reduction}
\label{sec:obs}
To characterize the internal kinematics of NGC 6440, we measured the
RVs of resolved, individual stars distributed over the entire radial
extension of the system by using a multi-instrument approach.

\begin{figure}[ht!]
\centering
\includegraphics[width=18cm, height=7.8cm]{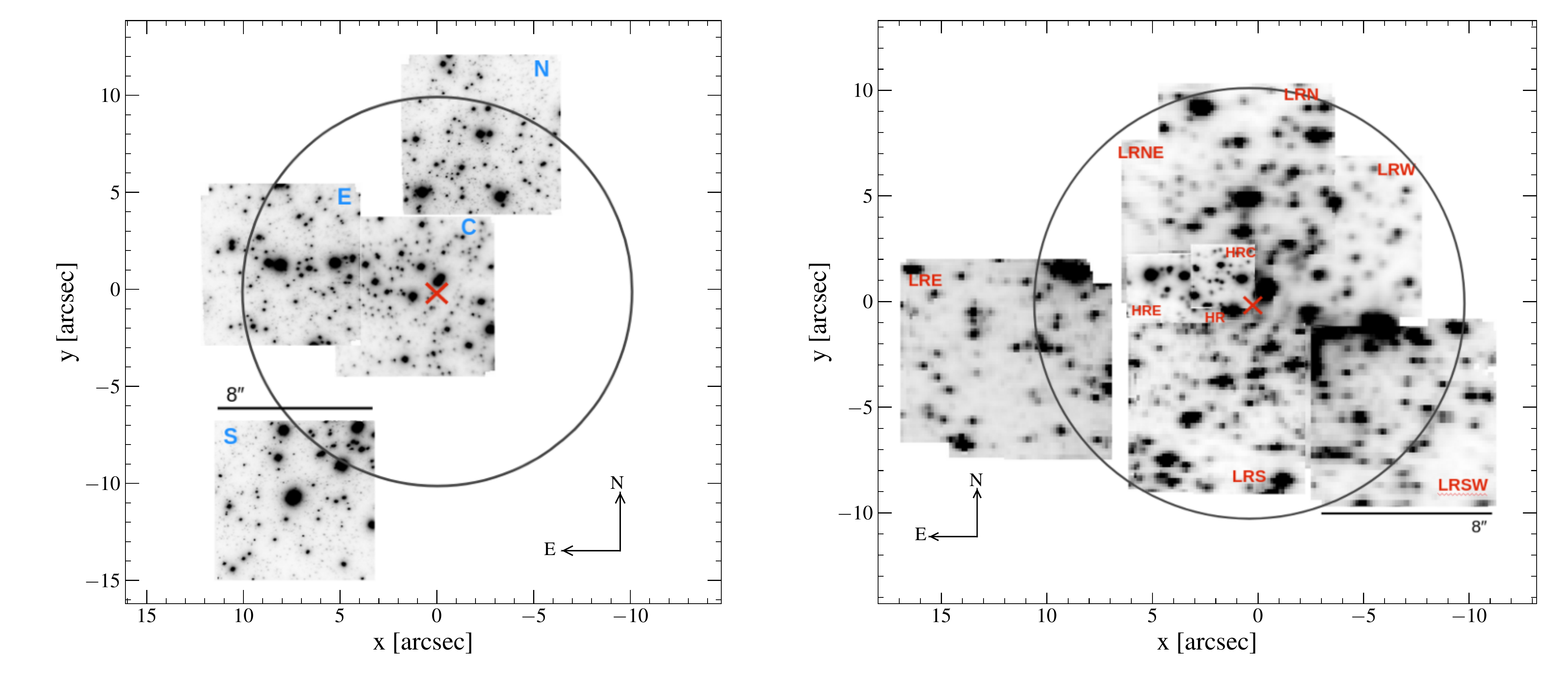}
\centering
\caption{{\it Left:} Reconstructed $I$-band images of the MUSE/NFM
  pointings. The circle is centered on the cluster center (red cross,
  from \citealp{pallanca+21}) and has a radius of $10\arcsec$.  {\it
    Right:} Reconstructed mosaic of the SINFONI/LR fields (each
  sampling $8\arcsec\times 8\arcsec$ on the sky) and SINFONI/HR
  pointings (with a $3\arcsec\times 3\arcsec$ field of view).  The red
  cross and the circle are as in the left panel.  In both panels the
  names of the pointings are labelled.}
\label{fig:mosaico}
\end{figure}

\begin{figure}[ht!]
\centering
\includegraphics[width=9.5cm, height=9.5cm]{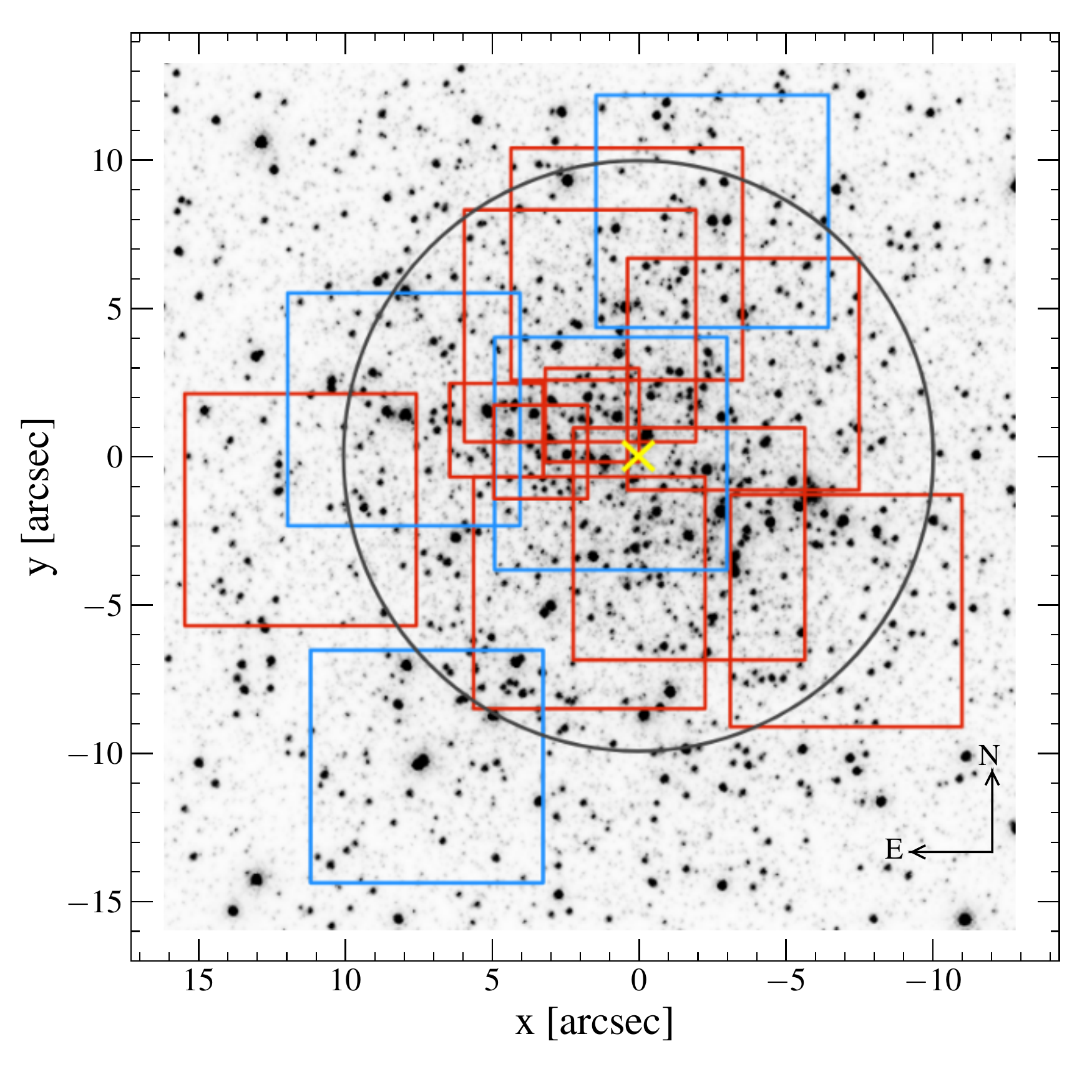}
\centering
\caption{HST/WFC3 image of the central region of NGC 6440 with the
  areas sampled by the spectroscopic observations overplotted: the
  blue squares show the fields of the four MUSE/NFM pointings ($\sim
  8\arcsec\times 8\arcsec$ on the sky), the red large squares are the
  SINFONI/LR fields ($\sim 8\arcsec\times 8\arcsec$), and the small red
  squares are the SINFONI/HR pointings ($\sim 3\arcsec\times
  3\arcsec$).  The black circle has a radius of $10\arcsec$ and is
  centered on the cluster center (yellow cross, from
  \citealp{pallanca+21}).}
\label{fig:pointings}
\end{figure}

\begin{figure}[ht!]
\centering
\includegraphics[width=12.2cm, height=9.1cm]{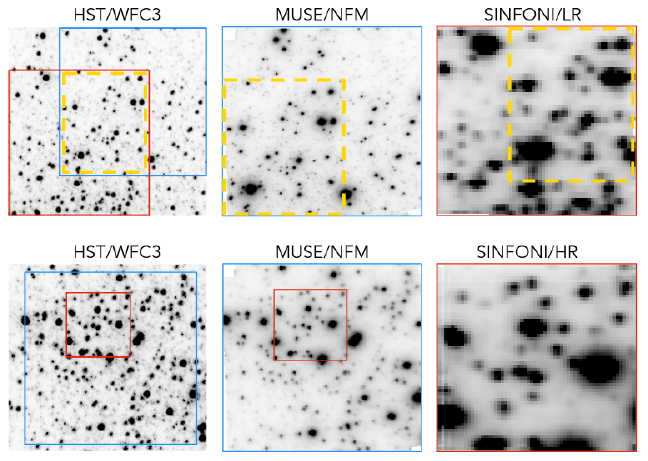}
\centering
\caption{\textit{Upper panels}: comparison among the HST/WFC3 image
  (left panel), the reconstructed $I$-band image of a MUSE/NFM
  pointing (central panel) and the stack image of a SINFONI/LR
  pointing (right panel), of a central area of NGC 6440. In the left
  panel, the blue square marks the MUSE/NFM field of view ($\sim
  8\arcsec\times 8\arcsec$ on the sky), as in the central panel, and
  the red region indicates the SINFONI/LR pointing ($\sim
  8\arcsec\times 8\arcsec$) shown in the left panel, while, in all the
  panels, the yellow dashed lines mark the common area among the three
  images. \textit{Bottom panels}: comparison among the images of the
  same area of the cluster 
  (bounded by the red square, $\sim 3\arcsec\times 3\arcsec$ on the sky)
  obtained using HST/WFC3 (left), MUSE/NFM (central), and
  the stack image of a SINFONI/HR pointing (right).
  The blue square in the left panel marks the area of MUSE/NFM field ($\sim
  8\arcsec\times 8\arcsec$ on the sky) shown in the central panel.}
\label{fig:confronto_instruments}
\end{figure}

\begin{itemize}
\item{\it MUSE -} The spectra of the stars in the innermost cluster
  regions were acquired with the AO-assisted integral-field
  spectrograph MUSE in the Narrow Field Mode (NFM) configuration
  \citep{bacon+10}, as part of the NFM science verification run
  (program ID: 60.A-9489(A), PI: Ferraro, see Table \ref{tab:data}).  MUSE is located on the Yepun (VLT-UT4)
  telescope at the ESO Paranal Observatory. It consists of 24
  identical Integral Field Units (IFUs) and it is available in two
  configurations, Wide Field Mode (WFM) and NFM, the latter providing
  a higher spatial resolution.  MUSE/NFM is equipped with the
  GALACSI-AO module \citep{Arsenault+08, Strobele+12} and covers a
  $7.5\arcsec\times 7.5\arcsec$ field of view with a spatial sampling
  of $0.025\arcsec$/pixel. The spectral range samples from $4800$
  \AA\ to $9300$ \AA\ with a resolving power R $\sim3000$ at
  $\lambda\sim8700$ \AA.  Our MUSE dataset consists of a mosaic of
  four MUSE/NFM pointings centered within $15 \arcsec$ from the
  cluster center \citep{pallanca+21}.  For each pointing, multiple
  exposures, usually three, were acquired with a small dithering
  pattern and a rotation offset of $90 \arcdeg$ between consecutive
  exposures, in order to remove possible systematic effects and
  resolution differences between the individual IFUs.  Each exposure
  has been acquired with an exposure time of 850 s, and the DIMM
  seeing during the observations ranged from $0.45\arcsec$ to
  $0.8\arcsec$.  The MUSE/NFM dataset was reduced by using the
  standard MUSE pipeline \citep{Weilbacher+20}. In a first step, the
  pipeline applies the bias subtraction, flat fielding, and wavelength
  calibration for each individual IFU, and, in a second step, it uses
  these pre-processed data of each IFU to perform the sky subtraction
  and flux and astrometric calibration.  Also, all the data are
  corrected for the heliocentric velocity.  Then, the data from all 24
  IFUs are combined into a single datacube. As the
  last step, the pipeline provides a final datacube by
  combining the datacubes of the multiple exposures of each pointing,
  taking into account the offsets and rotations among the different
  exposures.  The mosaic of the reconstructed $I$-band images from the
  stacking of MUSE datacubes is shown in the left panel of Figure
  \ref{fig:mosaico}. Each pointing is labeled with a name according to
  its position with respect to the cluster center (``C'', ``S'',
  ``E'', and ``N'' standing for central, southern, eastern, and
  northern pointing, respectively).

\item{\it SINFONI - } The MUSE/NFM data analysis in the 
  innermost cluster regions has been complemented using additional AO-assisted
  integral-field observations, performed with the spectrograph SINFONI
  \citep{Eisenhauer+03} at the ESO-VLT, in the near-infrared range
  $1.1 - 2.45 \mu$m. The observations were conducted under ESO
  proposals 093.D-0319(A), PI: Lanzoni, and ID:195.D-0750(A), PI:
  Ferraro (see Table \ref{tab:data}), by
  using the $K-$band grating, providing a spectral resolution R $\sim$
  4000 and sampling the $1.95 - 2.45 \ \mu$m wavelength range.  The
  dataset covers a region within $\sim16\arcsec$ from the cluster
  center, and consists of 7 pointings acquired by adopting the spatial
  scale of $0.25\arcsec$/spaxel corresponding to a field of view of
  $8\arcsec\times 8\arcsec$ (hereafter named ``LR'' for low
  resolution), and 3 pointings with the spatial scale of
  $0.1\arcsec$/spaxel and a $3\arcsec\times 3\arcsec$ field of view
  (hereafter, ``HR'' for high resolution).  Multiple exposures (usually 6 for
  the LR pointings, and 12 for the HR fields) of 20-30 s each were
  performed on the target and, for background subtraction purposes, on
  a sky position located $\sim 165\arcsec$ from the center, following
  the target-sky sky-target sequence.  The observations have been
  executed under an average DIMM seeing of $\sim 0.8\arcsec$, leading
  to a Strehl ratio between 10 and 40.  The data reduction was
  performed by using esorex (3.13.6) following the workflow 3.3.2
  under the EsoReflex environment \citep{Freudling+13}. The pipeline
  first corrects all target and sky exposures for darks, flats,
  geometrical distortions, and differential atmospheric
  refraction. Then, the sky background is subtracted by using the sky
  exposures, the wavelength calibration is performed through the
  observations of a Th-Ar reference arc lamp, and, the
  datacubes are built for each exposure by combining the corrected
  target frames.  The right panel of Figure \ref{fig:mosaico} shows
  the reconstructed image of the SINFONI pointings obtained by
  stacking the datacubes in the wavelength range $2.15 - 2.18
  \mu$m. Clearly, some stars are in common with the MUSE dataset, but
  the SINFONI pointings, in spite of a worse angular resolution, also
  sample the west and south-west regions around the cluster center
  that remained uncovered by MUSE (see Figure \ref{fig:pointings}).
In Figure \ref{fig:confronto_instruments} we compare the same cluster region as
seen in the reconstructed MUSE and SINFONI images (central and right
panels, respectively) and in the HST observations (left panels). This
well illustrates the exceptional resolving capabilities of the AO
systems used in the SINFONI (especially in the HR setput) and, even
more, in the MUSE observations, which are
mandatory to obtain large samples of RVs of individual stars in the
high-density core of dense stellar systems like NGC 6440.

\item{ KMOS -} To investigate the cluster kinematics at intermediate
  distances from the center, we have used the integral-field
  spectrograph KMOS \citep{Sharples+13} at ESO-VLT, which is equipped
  with 24 IFUs that can be allocated within a 7.2$\arcmin$ diameter
  field of view. Each IFU covers a projected area on the sky of about
  $2.8\arcsec \times 2.8\arcsec$, with a spatial sampling of
  $0.2\arcsec$/pixel. We have used the YJ grating covering the
  1.025-1.344 $\mu$m spectral range at a resolution R $\sim$ 3400,
  corresponding to a spectral sampling of $\sim 1.75$ \AA/pixel. The
  data have been collected as part of the ESO Large
  Program ID: 193.D-0232(A). These consist in 12 pointings within
  $\sim 6\arcmin$ from center, each one obtained with three
  sub-exposures 30 s long.  Every IFU is typically centered on one
  star, selected at $J < 14$ along the red giant branch (RGB) of the
  cluster from the near-infrared SOFI catalog available at the web
  site \url{http://www.bo.astro.it/~GC/ir_archive/}
  \citep{valenti+04,valenti+07}
and a 2MASS catalog in the $J$, $H$, and $Ks$ filters sampling the
outer regions.
The KMOS data reduction has been performed by using the dedicated
pipeline\footnote{\label{note1}http://www.eso.org/sci/software/pipelines/}
executing background subtraction, flat field correction, and
wavelength calibration.

\item{ FLAMES -} The external regions of NGC 6440 have been sampled
  out to $\sim 12\arcmin$ from the center by using the fiber-fed
  multi-object spectrograph FLAMES \citep{Pasquini+02} in the
  GIRAFFE/MEDUSA mode. This configuration consists of 132 fibers, each
  one with an aperture of 1.2$\arcsec$ on the sky, that can be
  allocated over a field of view of 25$\arcmin$ in diameter. As for
  KMOS, also the FLAMES observations have been performed within the
  MIKiS survey under the ESO Large Program 193.D-0232(B).  The spectra
  were acquired with the HR21 grating setup, which provides a
  resolving power $R\sim 18000$ between 8484 and 9001 \AA. Five
  pointings have been performed, each one securing two exposures of
  2700 s.  The targets are RGB stars brighter than $J = 14$ selected
  from the same photometric SOFI catalog used for the KMOS targets and
  from the 2MASS catalog.  The dataset has been reduced with the
  dedicated ESO pipelines\footnote{http://www.eso.org/sci/software/pipelines/}, including bias subtraction,
  flat-fielding correction, wavelength calibration, and extraction of
  one-dimensional spectra.  For each spectrum, the sky background has
  been subtracted, using a master sky spectrum obtained from the sky
  exposures acquired with 15-20 dedicated fibers in each pointing.
\end{itemize}

\begin{deluxetable*}{clcCC}
\tablecaption{Spectroscopic datasets for NGC 6440.}
\tablewidth{0pt}
\setlength{\tabcolsep}{10pt} 
\renewcommand{\arraystretch}{1} 
\tablehead{
\colhead{ Name }  &  \colhead{}  &
\colhead{Date} & \colhead{N$_{\rm exp}$} & \colhead{$t_{\rm exp}$ [s]}  }
\startdata
\hline
\multicolumn{5}{c}{MUSE/NFM}\\
\hline
C & & 2018-09-14 & 3 & 850\\
E & & 2018-09-14 & 3 & 850\\
N & & 2018-09-12 & 3 & 850\\
S & & 2018-09-13 & 2 & 850\\
\hline
\multicolumn{5}{c}{SINFONI}\\
\hline
HR  & & 2014-08-16 & 5 & 20\\
LRNE & & 2014-08-14 & 5 & 20\\
LRSE & & 2014-08-14 & 4 & 20\\
HRC & & 2015-06-24 & 6 & 30\\
   & & 2016-07-21  & 6 & 30\\
HRE & & 2015-06-25 & 6 & 20\\
   & & 2016-07-29  & 6  & 20\\
LRE & & 2016-07-30 & 6 & 30\\
LRN & & 2016-06-30 & 6 & 30\\
LRS & & 2016-07-30 & 6 & 30\\
LRSW & & 2015-07-17 & 6 & 30\\
LRW & & 2016-07-29 & 6 & 30\\
\hline
\multicolumn{5}{c}{KMOS}\\
\hline
kmos\TextUnderscore{}1 & & 2014-05-10 & 3 & 30 \\
kmos\TextUnderscore{}2 & & 2014-07-13 & 3 & 30 \\
kmos\TextUnderscore{}3 & & 2014-07-04 & 3 & 30 \\
kmos\TextUnderscore{}4 & & 2014-07-13 & 3 & 30 \\
kmos\TextUnderscore{}5 & & 2014-07-13 & 3 & 30 \\
kmos\TextUnderscore{}6 & & 2014-07-13 & 3 & 30 \\
kmos\TextUnderscore{}7 & & 2014-07-13 & 3 & 30 \\
kmos\TextUnderscore{}8 & & 2014-07-13 & 3 & 30 \\
kmos\TextUnderscore{}9 & & 2014-07-13 & 3 & 30 \\
kmos\TextUnderscore{}10 & & 2014-07-13 & 3 & 30 \\
kmos\TextUnderscore{}external\TextUnderscore{}1 & & 2014-07-20 & 3 & 30 \\
kmos\TextUnderscore{}external\TextUnderscore{}2 & & 2014-07-20 & 3 & 30 \\
\hline
\multicolumn{5}{c}{FLAMES}\\
\hline
flames\TextUnderscore{}1 & & 2014-06-19 & 2 & 2700 \\
flames\TextUnderscore{}2 & & 2014-06-20/23 & 2 & 2700 \\
flames\TextUnderscore{}3 & & 2014-06-23/2014-07-08 & 2 & 2700 \\
flames\TextUnderscore{}alt\TextUnderscore{}1 & & 2014-07-08 & 2 & 2700 \\
flames\TextUnderscore{}alt\TextUnderscore{}2 & & 2014-07-08 & 2 & 2700 \\
\hline
\enddata
\tablecomments{Name, observation date, number of exposures (N$_{\rm
    exp}$), and exposure time ($t_{\rm exp}$, in seconds) of each
  exposure for the MUSE/NFM, SINFONI, KMOS and FLAMES pointings
  analyzed in this paper.  In the name of the SINFONI pointings, LR
  and HR indicate the two different instrument configurations used
  here (low and high resolution, respectively).}
\label{tab:data}
\end{deluxetable*}


\section{Analysis}
\label{sec:analysis}
To properly analyze the four datasets, each one acquired with a
different instrument, we have performed specific analyses. They are
fully described in dedicated papers and briefly summarized below.

\subsection{MUSE/NFM dataset}\label{muse}
For the extraction of the spectra from the MUSE/NFM datacubes we used
the code PampelMuse presented in \citet{kamann+13}. This is a software
dedicated to the extraction of individual stellar spectra from MUSE
data in crowded regions of the sky, such as GC cores, by performing a
source deblending via wavelength-dependent point spread function (PSF)
fitting. While all the details can be found in \citet{kamann+13}, in
the following we briefly describe the main steps of the procedure.

Besides the spectroscopic datacube, PamelMuse needs in input a
photometric reference catalog providing the coordinates and magnitudes
of all the stars across the field of view.  To ensure high photometric
completeness and astrometric accuracy, we have used the HST/WFC3
catalog presented in \citet{pallanca+21}, and, to properly include
also, the brightest stars ($I<16$), which are saturated in those
long-exposures used to build the catalog above, we analyzed the HST/WFC3 images acquired under the
proposal GO15232 in the F555W and F814W filters (hereafter, $V$ and
$I$, respectively). The photometric analysis of this dataset has been
performed following the procedures described in
\citet{anderson_king_2006}, using the publicly available program
img2xym\TextUnderscore WFC.09x10.  To place the instrumental
coordinates onto the absolute astrometric system, and to calibrate the
instrumental magnitudes we have used the stars in common with the
catalog of \citet{pallanca+21}.
For source deblending purposes, PampelMuse also needs in input an
analytical PSF model.  We thus selected the MAOPPY function
\citep{fetick+19}, which is already implemented in the code
\citep[see][]{gottgens+21} and is designed to accurately reproduce the
typical double-component (core and halo) shape of the AO-corrected PSF
in MUSE/NFM observations.
Once setting the inputs, the spectra are extracted from the observed
datacubes through subsequent steps. First, a sub-sample of isolated
stars, on which the PSF will be modelled, is selected according to several
criteria, including: signal-to-noise ratio higher than a certain threshold (S/N$>5$),
relative contribution from neighboring sources negligible in region where the PSF is modelled, and no bright stars within a distance equal to 1.5 $\times$ the PSF definition radius.
The S/N is mainly
estimated from the magnitudes in the input catalog and an initial
guess on the PSF parameters. In the second step, the PSF fitting
procedure is applied to the selected sub-sample of stars in each
individual slice of the datacube, providing, as output, the wavelength
dependencies of the PSF parameters and of the source coordinates.
Finally, these wavelength dependencies are adopted in the PSF-fitting
procedure that is performed through the slices of the datacube, to
extract the spectra of all the sources present in the MUSE field of
view.  Among all the extracted spectra, we selected only those marked
as ``good'' by PampelMuse, which correspond to individual stars with
S/N $\geq 10$.

\begin{figure}[ht!]
\centering
\includegraphics[width=17cm, height=8.7cm]{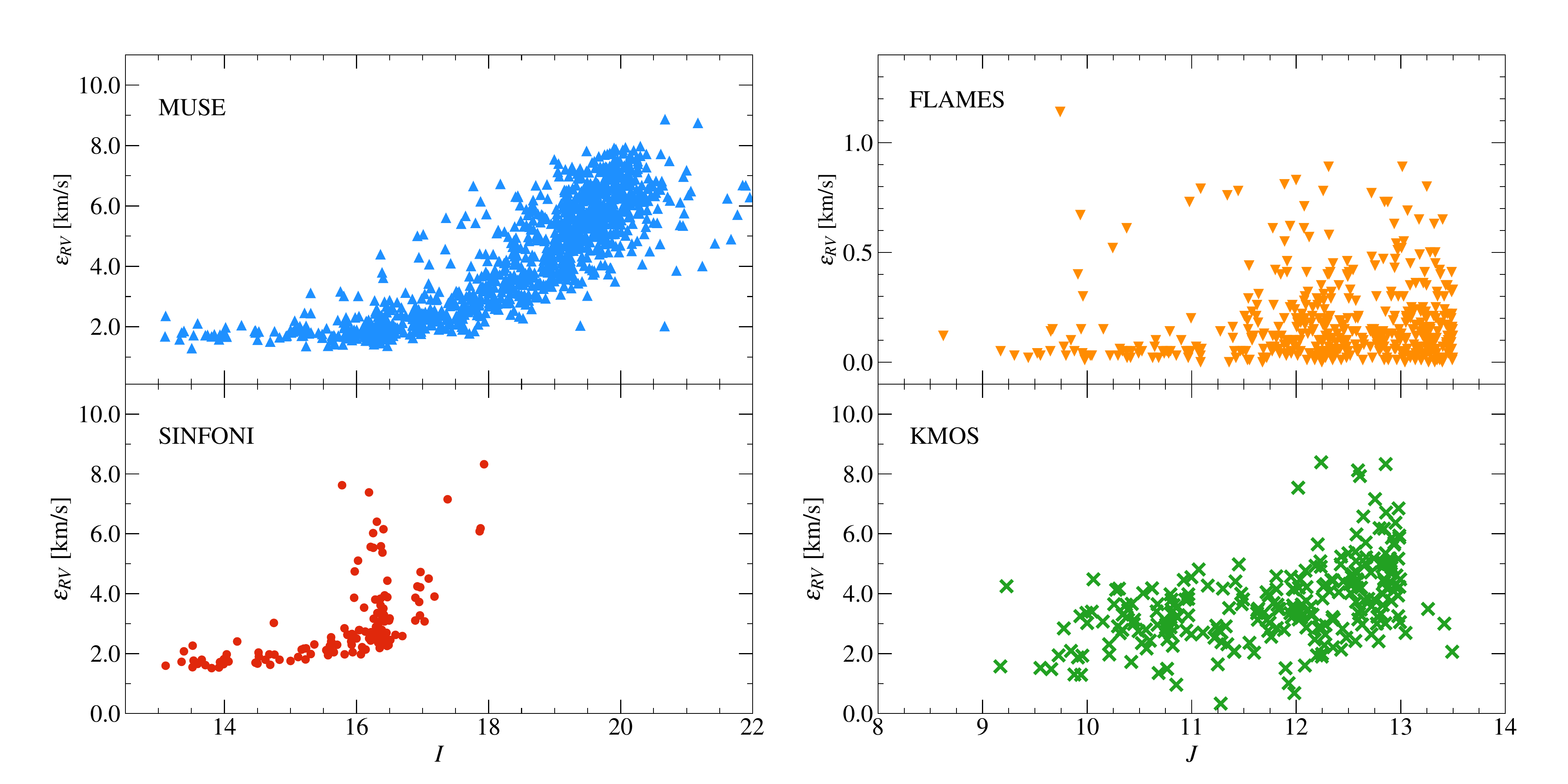}
\centering
\caption{RV uncertainty ($\epsilon_{\rm RV}$) as a function of the
  star magnitude for the observed targets in the MUSE/NFM, SINFONI,
  FLAMES, and KMOS samples (top-left, bottom-left, top-right, and
  bottom-right panels, respectively, see labels).}
\label{fig:err_mag}
\end{figure}

The RVs of the target stars have been measured from the Doppler shifts
of the Calcium Triplet lines in the wavelength range $8450-8750$ \AA.
To this end, the extracted spectra have been normalized to the
continuum (estimated through a spline fitting in the $7300 - 9300$ \AA
\ range).  Then, a library of template synthetic spectra has been
computed with the SYNTHE code (\citealt{Sbordone+04} and
\citealt{kurucz+05}), assuming an $\alpha-$enhanced chemical mixture
and the cluster metallicity ([$\alpha$/Fe]$=0.34$ dex and [Fe/H]$=-0.56$ dex;
\citealt{origlia+08}), and adopting a set of atmospheric parameters
(effective temperature and gravity) appropriate for the evolutionary
stage of the target stars, as derived from the CMD. The template
spectra have been convolved with a Gaussian profile to reproduce the
spectral resolution of MUSE, and re-sampled at the same pixel size of
the observed spectra.  The procedure adopted to measure the target RV
computes the residuals between the observation and each template
spectrum of the library shifted in velocity in steps of 0.1 km
s$^{-1}$. The distribution of the residuals showing the smallest
standard deviation ($\sigma_{\rm min}$) provides the best-fit
synthetic spectrum (hence, the best estimate of the stellar
atmospheric parameters), and from the minimum of this distribution the
RV of the target is derived.  A value of S/N independent of that
obtained by PampelMuse has been computed for each spectrum as the
ratio between the average of the counts and their
standard deviation in the wavelength range $8000 - 9000$ \AA.  We will
use this S/N estimate in the following analysis.\\
The uncertainties on the RV measures have been estimated by means of
Monte Carlo simulations.  By adding different amounts of Poisson noise
to the adopted synthetic templates, we simulated $\sim 9000$ observed
spectra with S/N ratios ranging from 10 to 90, in step of 10, running
100 simulations for every considered value of S/N.  Then, this sample
has been analyzed as for real observations, computing the residuals
between the simulations and each synthetic spectrum of the library
progressively shifted in velocity, as described above.  For each
synthetic spectrum we selected the distribution of residuals showing
the smallest standard deviation ($\sigma_{\rm min}$) and we adopted
the corresponding value of RV.  This allowed us to plot the difference
between the output and the input RVs ($\Delta$RV) as a function of
$\sigma_{\rm min}$, from which a polynomial relation between the two
parameters has been obtained, with the values of $\Delta$RV
increasing for increasing $\sigma_{\rm min}$ (hence, for decreasing
S/N ratios).  Finally, knowing the value of $\sigma_{\rm min}$ of the
observed spectra, we used this relation to determine the corresponding
value of $\Delta$RV, which has been adopted as RV uncertainty.  The
typical RV errors are $\sim 2$ km s$^{-1}$ for the brightest stars
($I<16$), and they increase up to $\sim 8$ km s$^{-1}$ for the
faintest targets, as shown in the top-left panel of Figure
\ref{fig:err_mag}.  To check that the measured RVs were homogeneous
among the different MUSE pointings, we compared the values obtained
for the stars in common between the two overlapping fields (the
central and the east ones), and the average RV value obtained in each
pointing, always finding a good agreement within the errors.  In the
case of multiple exposures for the same star, we adopted as the final
RV the weighted mean of all the measures, by using the individual
errors as weights.

The final MUSE catalog consists of 1128 individual RV measures for
stars located between $\sim 0.1\arcsec$ and $\sim 18\arcsec$ from the
cluster center, in the magnitude range $13< I < 22$ 
(see Table \ref{tab:catalogo}).
The position of the stars on the plane of the sky is
shown in the right panel of Figure \ref{fig:mappa} (blue triangles), while the first panel on the left of Figure \ref{fig:cmd} shows the corresponding CMD.

\subsection{SINFONI dataset}
\label{sec:sinfo}
A forthcoming paper (Cristina Pallanca et al., 2023, in preparation)
will be specifically devoted to the detailed description of the
procedure adopted for the analysis of the SINFONI spectra. Here we
just summarize the main key points.

First of all, in each observed datacube, we selected all the spaxels
with photon counts above a threshold of $10 \sigma$ the background
level. Setting the threshold level to such a high value guantantees
the selection of only the spaxels acquired at the largest S/N ratio.
A value of RV has been measured from the 1D spectrum extracted 
from each selected spaxel,
applying a procedure analogous to that adopted
for the MUSE/NFM data, using the Doppler shift of the $^{12}C^{16}O$
band-heads, instead of the Calcium Triplet lines.  More specifically,
the observed spectra have been compared with synthetic templates
progressively shifted in velocity, computed with the SYNTHE code
\citep{Sbordone+04, kurucz+05} in the appropriate NIR wavelength range
and at the SINFONI spectral resolution.  We used synthetic spectra
computed for 10 pairs of effective temperature and surface gravity
properly sampling the entire RGB of the cluster, with iron and
$\alpha-$element abundances measured by \citet{origlia+08},
plus 7 additional models with appropriate carbon-depletion [C/Fe]$=-0.36$ dex 
\citep{origlia+08} for stars above the RGB bump. This is to
take into account the fact that the deepness of the CO band-heads
depends on both the temperature and the chemical abundance, and the
stars above the RGB bump could be depleted in carbon. The best
estimates of the RV and its uncertainty have been evaluated as in the
case of the MUSE data, from the distribution of the residuals between
the observed and the synthetic spectra showing the smallest standard
deviation, and from Monte Carlo simulations, respectively.  The RV
errors are of the order of 2 km s$^{-1}$ and their trend as a function
of the star magnitude is shown in the bottom-left panel of Figure
\ref{fig:err_mag}.

From the cross-correlation with the HST catalog described above we
finally identified the resolved stars in each SINFONI datacube. 
To enhance the quality of the dataset, for each source
we used only the RV measured from the central spaxel (with
the largest S/N ratio). By comparing the values obtained from
repeated observations of the same stars (which are available
especially in the HR fields), and the average RVs of each field, we
verified that no relevant offsets in the RV zero points are present,
the average differences being about 2 km s$^{-1}$, which is consistent 
with the value of the standard deviation. 
Hence, the final RV catalog has been generated by
averaging the values measured in different datacubes in case of multiple
exposures of the same star, using the estimated errors as weights. 
If a star was sampled both in an HR and in a LR pointing, we
kept the value measured in the former for the higher spatial resolution 
of this instrumental setup. 

As discussed below, to reliably investigate the internal kinematics of
NGC 6440 we restricted the sample of RVs to the safest measures only.
The SINFONI data sample the cluster core where stellar crowding is
critical, but the procedure used to extract the spectra includes no
source deblending algorithms.  Hence, the derived RVs might be
affected by the presence of brighter neighboring stars, and this can
impact the final results in terms of the cluster velocity dispersion
and systemic rotation.  To address this issue and select only the
spectra contributed by the light of individual stars, we applied the
procedure described in \citet{leanza+22}, which is briefly summarized
here.  Using as inputs the list of detected stars from the HST catalog
and the PSF model adopted in the SINFONI data reduction, the procedure
computes the contamination parameter ($C$) as the ratio between the
fraction of light contributed by the first contaminant and that of the
target under analysis, where the first contaminant is the neighboring
source providing the second largest contribution of light to the central
spaxel, after the target itself. 

For the final SINFONI sample we selected only
the safest targets, with negligible contamination from neighboring sources, by including only the stars with contamination parameter lower than 3\% ($C<0.03$).
The final catalog consists of
138 RVs for individual stars located between $1.1\arcsec$ and
$15.0\arcsec$ from the cluster center (red circles in the right panel of Figure \ref{fig:mappa}), in the magnitude range $13< I <
18$ (see Table \ref{tab:catalogo}). 
The CMD of the targets is shown in the second panel from the left of
Figure \ref{fig:cmd}.
\begin{deluxetable*}{cCCC}
\tablecaption{Summary of the different datasets used in this work.}
\tablewidth{0pt}
\setlength{\tabcolsep}{10pt} 
\renewcommand{\arraystretch}{1} 
\tablehead{
\colhead{ Dataset } & \colhead{ Number of stars }  & \colhead{Radial region in arcsec}  &  \colhead{Magnitude range}}
\startdata
MUSE/NFM & 1128 & 0.1 - 18.0 & 13.0<I<22.0 \\
SINFONI & 138 & 1.1 - 15.0 & 13.0 < I < 18.0 \\
KMOS & 258 & 1.1 - 407.3 &  9.2< J < 13.5 \\
FLAMES & 448  & 22.0 - 778.5 & 8.6 < J < 13.5 \\
\enddata
\tablecomments{For each dataset, the columns list the name
of the instrument, the number of stars with RV measured, 
the sampled radial region expressed as the distance from 
the center in arcsec, and the magnitude range of the target stars.}
\label{tab:catalogo}
\end{deluxetable*}
\subsection{KMOS and FLAMES datasets}
\label{sec:kmos_flames}
\begin{figure}[ht!]
\centering
\includegraphics[width=17.6cm, height=8.4cm]{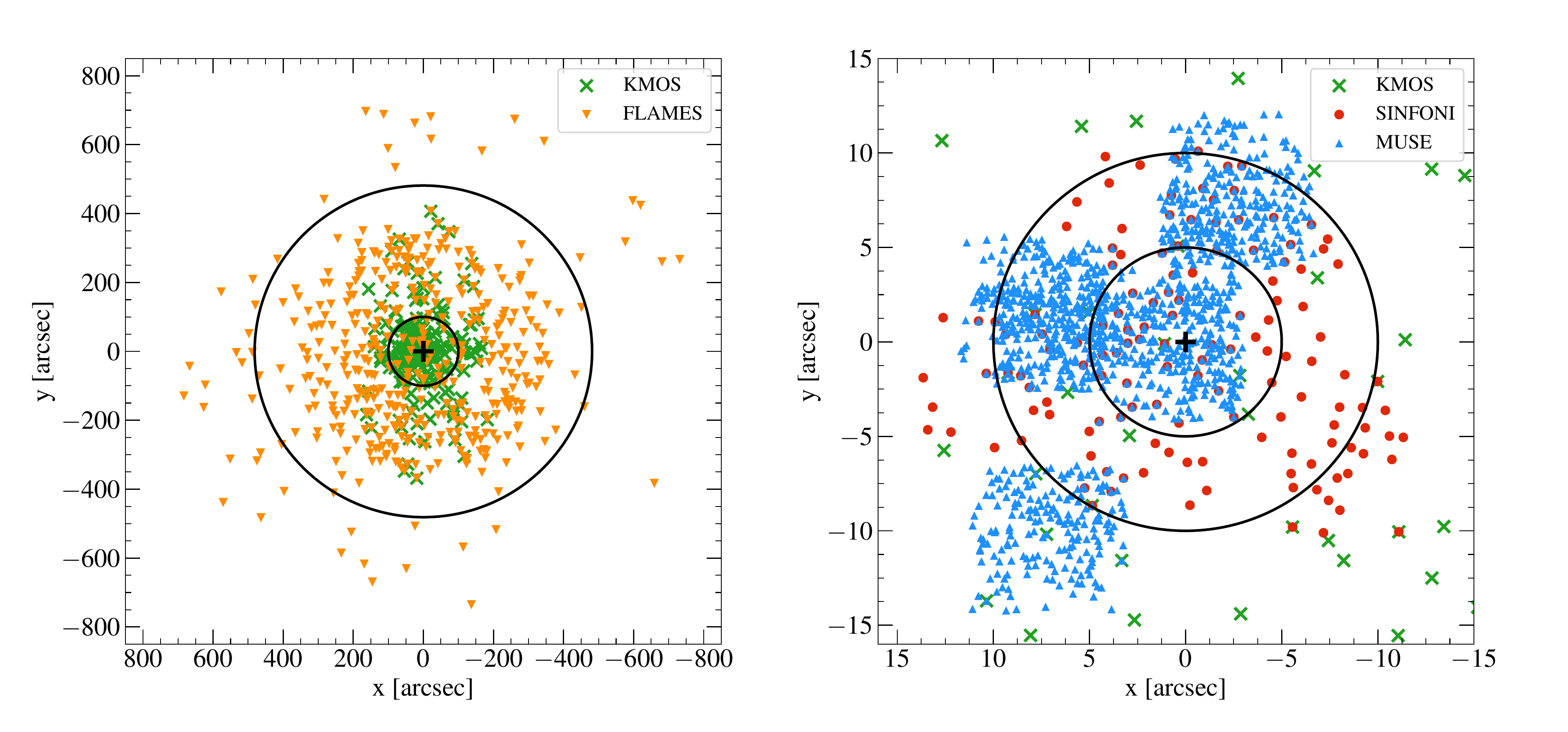}
\centering
\caption{Maps on the plane of the sky, with respect to the adopted
  cluster center (black cross), of the stars with measured RV for each dataset. The left panel shows the external portion of
  the cluster sampled by the FLAMES (orange triangles) and the KMOS
  (green crosses) datasets. The two circles mark distances of
  100$\arcsec$ and 481.4$\arcsec$ (corresponding to the truncation
  radius of the cluster, see \citealt{pallanca+21}) 
  from the center. The right panel is
  focused on the central region covered by the MUSE (blue triangles) and
  SINFONI (red circles) samples. A few KMOS targets are also visible
  (green cross). The two circles mark distances of 5$\arcsec$ and
  $10\arcsec$ from the center.}
\label{fig:mappa}
\end{figure}

\begin{figure}[ht!]
\centering
\includegraphics[width=17cm, height=10.2cm]{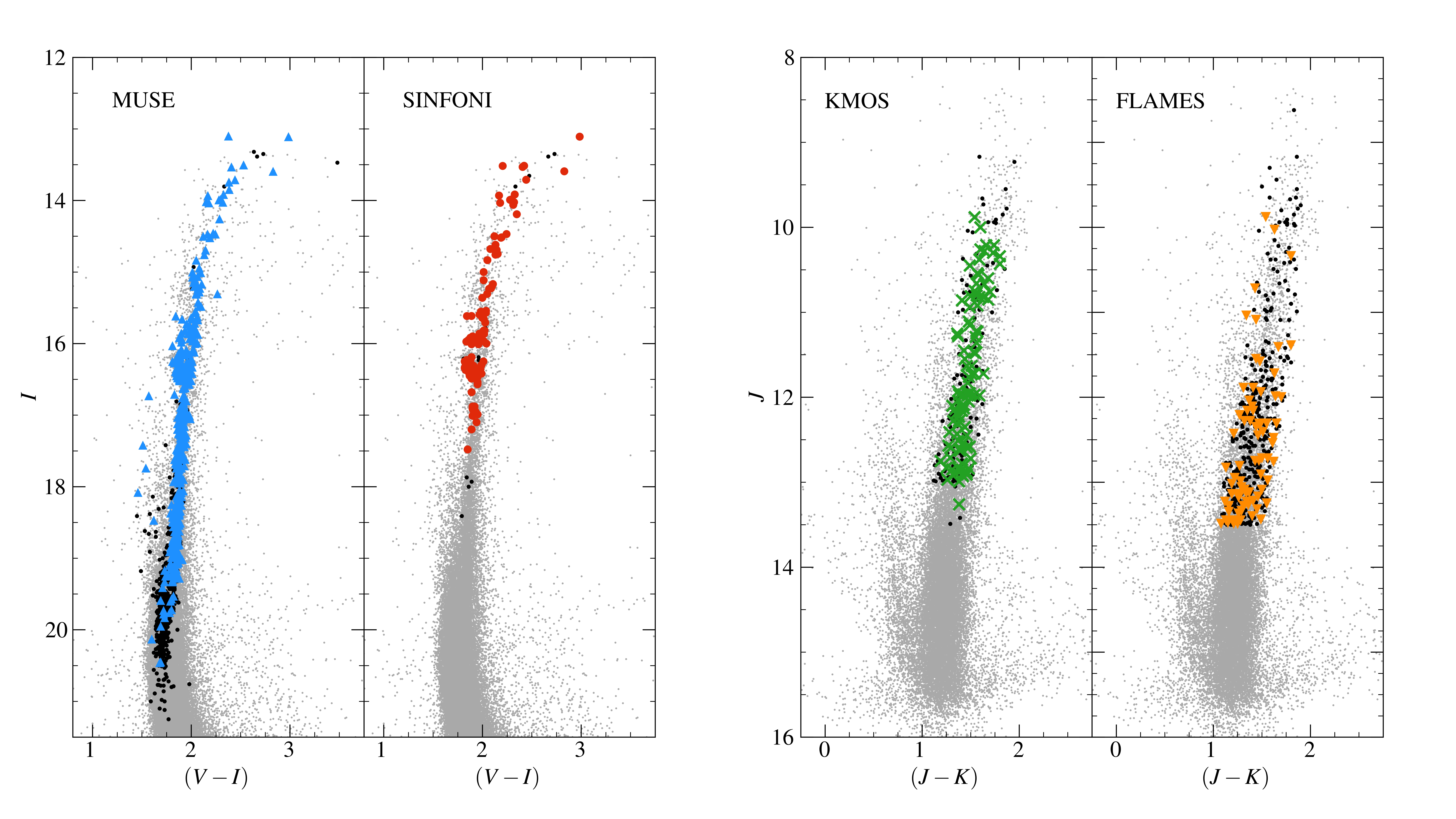}
\centering
\caption{CMDs of NGC 6440 with the star having measured RV
  highlighted. \textit{Left panels}: the gray dots show the ($I, V-I$)
  CMD obtained from the photometric catalog discussed in
  \citet{pallanca+21}, with overplotted the MUSE (left) and SINFONI
  (right) samples. The blue triangles on the left and the red circles
  on the right mark, respectively, the MUSE and SINFONI targets after
  all the membership and quality selections described in Sections
  \ref{membership} and \ref{sec:results}, respectively,
  while the black dots are the rejected targets.
  \textit{Right panels}: ($J, J-K$) CMD obtained from the reference
  SOFI/2MASS catalog (gray dots in both the panels; see Section
  \ref{sec:obs}), with the KMOS and FLAMES datasets highlighted on the
  left and on the right, respectively. The green crosses on the left
  and the orange triangles on the right correspond to the KMOS and
  FLAMES samples selected for the kinematic analysis, respectively
  (see Section \ref{membership} and \ref{sec:results}). The black dots indicate the
  rejected stars. Note that in each panel, the targets observed 
  by two or more instruments are also included.}
\label{fig:cmd}
\end{figure}

The procedures adopted to measure the RVs of the KMOS and FLAMES
targets are fully described in \citet{ferraro+18a,ferraro+18b}, where
the MIKis survey is presented.  Briefly, for the KMOS observations,
the 1D spectra have been extracted manually by visually inspecting
each IFU and selecting the most exposed spaxel, which corresponds to
the stellar centroid.  Then, after correction for heliocentric
velocity, both KMOS and FLAMES spectra have been cross-correlated with
template spectra following the procedure described in
\citet{tonry_davis_79}, which is implemented in the FXCOR task under
the software IRAF.  
To verify that using two different methods does not introduce 
systematic effects in the RV measurements, 
we applied the cross correlation of IRAF to the MUSE spectra, 
obtaining results in perfect agreement with those 
obtained with the method described in Section \ref{muse}.
As in the cases of the other datasets, the
template spectra have been computed with the SYNTHE code
\citep{Sbordone+04, kurucz+05} in the appropriate wavelength range,
adopting the cluster metallicity and RGB atmospheric parameters, and
applying a convolution with a Gaussian profile to reproduce the
instrument spectral resolutions.  For KMOS observations, the RV has
been obtained from the cross-correlation with individual near-IR
features in the sampled wavelength range, and the RV uncertainties
have been derived using Monte Carlo simulations similar to those used
for MUSE and SINFONI.  For the FLAMES targets, the RV has been
measured in three different regions of the spectrum, each including a
large number of atomic lines, and the final value and its uncertainty
have been obtained, respectively, as the average of the three
measures, and their dispersion divided by the square root of 3.  The
typical RV errors are of the order of 1-5 km s$^{-1}$ for the KMOS
targets, while they decrease to $\sim 0.1$-0.3 km s$^{-1}$ for the
FLAMES measures (bottom-right and the top-right panels of Figure
\ref{fig:err_mag}, respectively). The final KMOS and FLAMES samples
consist of 258 and 448 RV measures, respectively 
(see Table \ref{tab:catalogo}).
The left panel of Figure \ref{fig:mappa} shows the position of the stars
on the plane of the sky with orange triangles and green crosses 
for the FLAMES and KMOS samples, respectively, 
while the corresponding CMDs are shown in the third (KMOS) and fourth panels (FLMAES) from the left of Figure \ref{fig:cmd}.

\subsection{Final combined catalog}\label{comb_catalog}
Before combining all the RV measurements in a single final catalog, we
have checked for possible systematic offsets among the different
catalogs.  These could be due to the different instrumental setups,
including the wavelength range used to derive the RVs, and also the
different spectral resolution of each spectrograph.

To this aim, we compared the RVs of the stars in common between two
datasets, using only reliable RV measures (i.e., high S/N and small RV
uncertainty) and adopting the values obtained from the FLAMES
observations as reference, due to the highest spectral resolution of
this instrument.  From the stars in common between FLAMES and KMOS, we
found an average offset of $-5.2$ km s$^{-1}$, which was then applied
to the KMOS measures for realigning this sample with the reference
one.  Unfortunately, no stars are in common between the FLAMES dataset
and the MUSE and SINFONI ones, since they sample very different
regions of the cluster.  Moreover, no reliably enough RV measures from
the KMOS observations have been found in common with the MUSE and
SINFONI datasets. Therefore, to realign the innermost samples with
the reference catalog, we compared, after excluding the obvious
outliers, the average velocities obtained from the FLAMES and the
MUSE catalogs finding a good agreement within the errors 
($-67.5 \pm 1.1$ km s$^{-1}$ and $-67.7 \pm 0.5$ km s$^{-1}$, respectively).
As last
step, a very small residual offset of $-1.0$ km s$^{-1}$ was detected
between the RVs of the stars in common between the MUSE and the
SINFONI datasets.  Hence, this offset was applied to the SINFONI RVs
to realign this sample with all other catalogs.

To create the final catalog, we combined the four datasets (summarize in Table 2) by performing a weighted mean of the RV measures by using the individual errors as weights, for the targets observed by more than one instrument.
The final catalog consists of 1831 RV measurements of individual 
stars \footnote{The final catalog including the identification number, right ascension, declination, RV measure and its error for each star is available for free download at: 
\url{http://www.cosmic-lab.eu/Cosmic-Lab/MIKiS_Survey.html}}
sampling the entire radial extension of the cluster, from
$0.1\arcsec$, out to $778\arcsec$ (corresponding to $\sim 1.6$ times
the truncation radius $r_t=481.4\arcsec$, and $\sim 15$ times
the three-dimensional half-mass radius $r_h=50.2\arcsec$; see
\citealt{pallanca+21}) from the cluster center, as shown in Figure \ref{fig:mappa}, and covering a wide
magnitude range ($13<I<22$, see Figure \ref{fig:cmd}).
  
\subsection{Cluster membership}
\label{membership}
\begin{figure}[ht!]
\centering
\includegraphics[width=14.5cm, height=10.5cm]{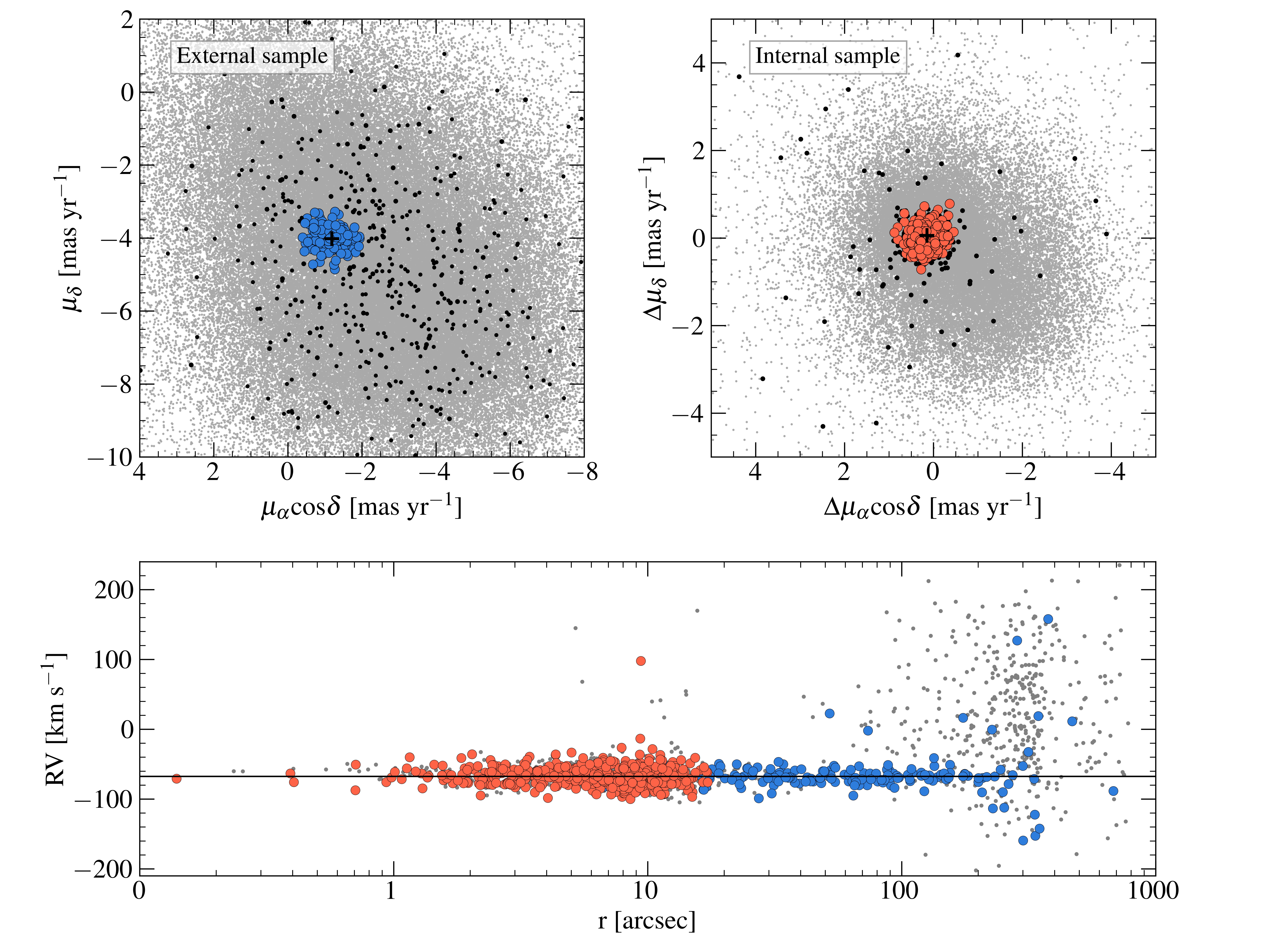}
\centering
\caption{\textit{Top-left panel}: VPD of the Gaia EDR3 dataset (gray
  dots) with the targets of the external sample selected as member stars
  marked with blue circles.
  The black dots show the targets rejected as field
  stars. The black cross marks the absolute motion of NGC 6440
  \citep{vasiliev+21}.  
  {\it Top-right panel}: VPD of the relative HST
  PMs obtained in \citet[][gray dots]{pallanca+19}. 
  The red circles mark the member stars selected from the internal sample, 
  while the black circles are the targets considered as non-member stars.
  {\it Bottom panel}: RVs of the final catalog
  as a function of the distance from the cluster center. 
  The large circles show the targets selected as cluster members, 
  color-coded as in the top panels. The targets rejected
  as field stars are marked with gray circles.}
\label{fig:members}
\end{figure}

Being a GC in the bulge direction, the contamination from field stars
of the population of NGC 6440 is not negligible. For this reason, a
thoughtful and accurate analysis was performed to properly address the
issue of cluster membership of the measured stars. For the external sample (FLAMES and KMOS), we took advantage of the proper motions (PMs) provided
in the Gaia EDR3 \citep{gaiaEDR3}: we selected as cluster members the
stars with PMs within 0.9 mas yr$^{-1}$ from the absolute motion of
NGC 6440 \citep{vasiliev+21} in the vector-point diagram (VPD, see the
top-left panel of Figure \ref{fig:members}), this value corresponding
to $\sim 3$ times the central velocity dispersion of NGC 6440 (see
Section \ref{sec:2ndv}), assuming a distance of $8.3$ kpc
\citep{pallanca+21}.  
The same could not be done for the internal sample 
(mostly MUSE and SINFONI)
because they either have no measured PM, or the
measures are not reliable, due to the limited capabilities of Gaia in
the very central regions of dense GCs like NGC 6440.  Therefore, to
identify the cluster members in the innermost samples, 
we used the same criteria based on the relative HST PMs
presented in \citet{pallanca+19}. The VPD and
the member selection of the internal sample are shown in the
top-right panel of Figure \ref{fig:members}.  The bottom panel of the
figure shows the measured RVs as a function of the distance from the
center, with the PM-selected member stars highlighted as large,
colored circles.  As apparent, the RVs of the bulk of cluster members
are centered at about $-67$ km s$^{-1}$, while field stars have
significantly different (especially, larger) RVs and become dominant
in the most external regions.  The residual field contamination that
appears to be still present in the PM-selected sample, especially at
large distances from the center, will be easily removed in the
following analysis by means of a $\sigma$-clipping procedure aimed at
excluding the obvious outliers (see Section \ref{sec:results}).

\section{Results}
\label{sec:results}
To properly explore the internal kinematics of NGC 6440, in the
following analysis we have used only stars with the most reliable RV
measures, selected by adopting the following criteria. Among
the PM-selected cluster members, only the targets with S/N $>15$ and
RV error $<5$ km s$^{-1}$ have been considered.
After applying these selections to the targets of the combined catalog 
(see Section \ref{comb_catalog}), we obtained a total sample of 704
targets. Figure \ref{fig:cmd} shows the positions of
the selected targets in the appropriate CMDs.  This is the sample of
RV measurements that we used to determine the systemic velocity and
the velocity dispersion profile, and to detect possible signatures of
rotation.

\subsection{Systemic velocity}
\label{sec:vsys}
\begin{figure}[ht!]
\centering
\includegraphics[width=16.5cm, height=6.2cm]{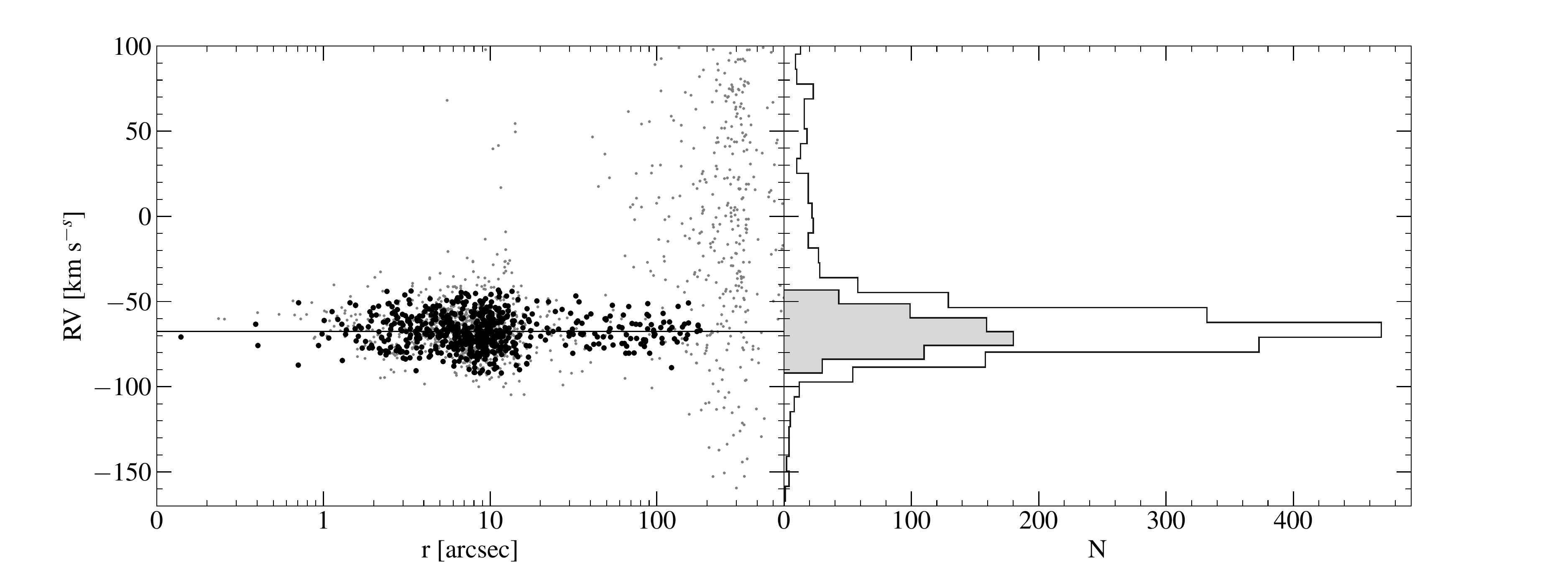}
\centering
\caption{{\it Left panel:} RVs of the final catalog (gray dots) as a
  function of the distance from the cluster center, with the 625 RVs
  used to determine the cluster systemic velocity (solid line)
  highlighted as black circles. {\it Right panel:} Number distributions
  of the final catalog (empty histogram) and of the targets used for
  the determination of $V_{\rm sys}$ (gray histogram).}
\label{fig:rv_dist}
\end{figure}

The measured RVs as a function of the distance from the center are
plotted in the left panel of Figure \ref{fig:rv_dist}, and the
corresponding RV distribution is shown in the right panel. The gray
dots and empty histogram refer to the entire final catalog
(1831 RV measures),
and the apparent well defined peak indicates the cluster systemic velocity
($V_{\rm sys}$).  
For the measure of $V_{\rm sys}$ only, in order to minimize the 
risk of a residual contamination
from field stars, from the confident sample of stars selected as described 
above, we considered only those with distances within $200\arcsec$ ($\sim 4 \times r_h$) from the center and with RVs in the range $-92$ km s$^{-1}<$RV$<-42$ km s$^{-1}$, 
and we applied a $3\sigma$-clipping procedure to remove the
obvious outliers. The resulting sample of 625 RVs is shown in the
left panel of Figure \ref{fig:rv_dist} as black circles, while its
distribution is plotted as a gray histogram in the right panel.
Hence, under the assumption that the RV distribution of the selected
stars is Gaussian, the value of $V_{\rm sys}$ and its uncertainty have
been computed through a Maximum-Likelihood method
\citep{Walker+06}. We obtained $V_{\rm \rm sys} = -67.5 \pm 0.4$ km
s$^{-1}$.  This estimate is in good agreement with the previous result
published in \citet[][$-67.8 \pm 1.0$ km s$^{-1}$]{Baumgardt+18},
while it disagrees with the value quoted in \citet[][$-76.6 \pm 2.7$
  km s$^{-1}$]{harris+96}.  In the following, we will indicate as $V_r
= {\rm RV} - V_{\rm sys}$ the RVs referred to the cluster systemic
velocity.

\subsection{Second velocity moment profile}
\label{sec:2ndv}
As discussed in previous papers \citep[e.g.,][]{lanzoni+18a,leanza+22}
the second velocity moment profile ($\sigma_{II}(r)$) derived from the
measured RVs represents a very good approximation of the projected
velocity dispersion profile $\sigma_P(r)$, in the case of no or
negligible systemic rotation, according to the relation:
\begin{equation}
  \sigma^2_P(r) = \sigma^2_{II}(r) - A^2_{\rm rot}(r),
\end{equation}
where $A_{\rm rot}$ is the amplitude of the rotation curve. For this
cluster no evident signatures of rotation have been clearly detected
in previous works. Therefore, we first compute the second velocity
moment profile to compare it with the previous results, and then, in
Section \ref{sec_vrot}, we complete the kinematic analysis of the
cluster investigating the possible presence  of internal rotation.

To determine the second velocity moment profile,
starting from the RVs sample selected as described is Section \ref{sec:results}, we have adopted the
standard approach already used in previous works (see
\citealt{ferraro+18a, lanzoni+18a, lanzoni+18b, leanza+22}): the RV
sample is divided into radial bins using a set of concentric annuli,
which are selected at increasing distance from the cluster center and
provide a good compromise between fine radial sampling and
statistically significant numbers of stars (at least 30) in each bin.
A 3$\sigma-$clipping procedure is performed on the RVs in each radial
bin to exclude the obvious outliers from the analysis. Then, we
applied a Maximum-Likelihood method \citep[][see also
  \citealp{Martin_2007}; \citealt{sollima+09}]{Walker+06} to compute
the dispersion of the $V_r$ values of the selected stars in each
bin. The uncertainties are estimated following the procedure described
in \citet[][]{pryor+93}.

\begin{figure}[ht!]
\centering
\includegraphics[width=12cm, height=10cm]{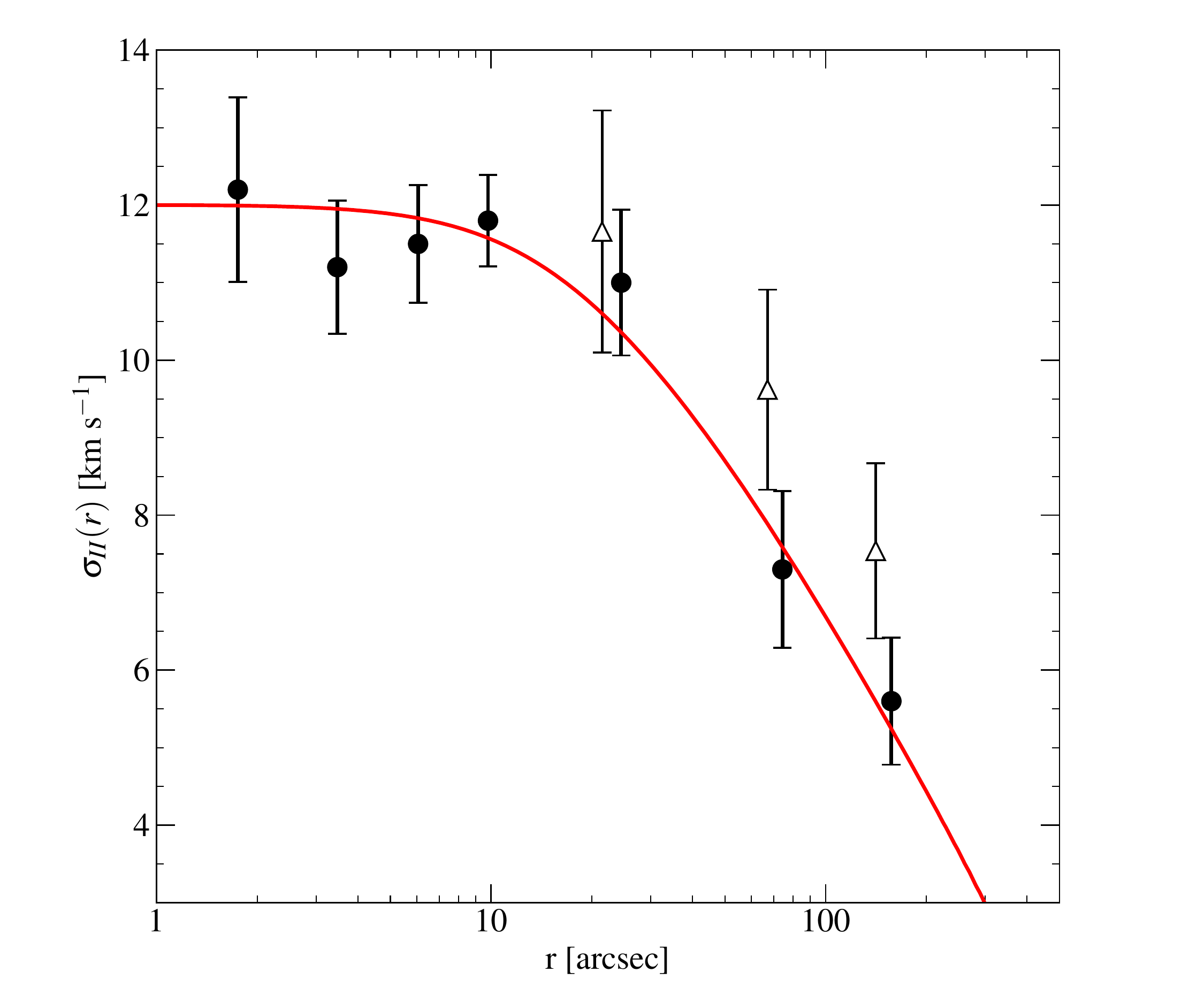}
\centering
\caption{Second velocity moment profile of NGC 6440 obtained from the
  measured individual RVs (solid circles).  The empty triangles show
  the profile derived by \citet{Baumgardt+18}. The red solid line
  represents the King model that best fits the star density profile of
  the cluster \citep{pallanca+21}.}
\label{fig:vdp}
\end{figure}

The resulting $\sigma_{II}(r)$ profile of NGC 6440 is shown in Figure
\ref{fig:vdp} (black circles) and listed in Table \ref{tab:2ndv_01}.
It nicely follows the King model (red line) that best fits the
observed star density profile of the cluster \citep{pallanca+21}.
We have estimated the central velocity
dispersion as the value that minimizes the residuals
between the observed velocity dispersion profile and the adopted King model
finding $\sigma_0 = 12.0 \pm 0.4$ km s $^{-1}$. 
The $1\sigma$ uncertainty has been obtained from the
solutions of the $\chi^2$ test for which $\chi^2 = \chi^2_{min} \pm 1$.

In Figure \ref{fig:vdp} we also compare our result with the 
observed profile obtained by \citet[][empty
triangles]{Baumgardt+18} from RV measures at intermediate and large
radii from the center. 
Formally, the two outermost points of \citet{Baumgardt+18} are larger
than ours, possibly due to a different membership selection applied 
in the two works, or an effect of residual field star contamination in the former.
The difference, however, is not significant (the error bars are just $1\sigma$),
but we verified that the King model that best fits the observed density
distribution would be unable to reproduce a velocity dispersion profile obtained
by combining our innermost four points and the three measures
by \citet{Baumgardt+18}, while it is well consistent with the 
determination provided in this work (solid circles and red line in Figure \ref{fig:vdp}).

\begin{deluxetable*}{RRRRCC}
\tablecaption{Second velocity moment profiles obtained for NGC 6440.}
\tablewidth{0pt}
\tablehead{
\colhead{ $r_i$ } & \colhead{ $r_e$ }  & \colhead{$r_m$}  &
\colhead{$N$} & \colhead{$\sigma_{II}$} & \colhead{$\epsilon_{\sigma_{II}}$} \\  
\colhead{ [arcsec] } & \colhead{ [arcsec] }  & \colhead{[arcsec]}  &  \colhead{ } &\colhead{km s$^{-1}$ }  &
\colhead{ km s$^{-1}$} 
}
\startdata
0.01  &  2.50  &  1.75  & 58 &  12.20 & 1.19  \\
2.50  &  4.50  &  3.47  & 96 &  11.20 & 0.86  \\
4.50  &  7.50  &  6.05  & 130 & 11.50 & 0.76  \\
7.50  &  13.00  & 9.79  & 228 & 11.80 & 0.59  \\
13.00 &  50.00  & 24.49 & 79  & 11.00 & 0.94  \\
50.00 & 100.00  & 74.23 & 37  & 7.30  & 1.01  \\
100.00 & 250.00 & 157.22 & 30 & 5.60  & 0.82  \\
\enddata
\tablecomments{The first three columns list the internal, external,
  and mean radii of each adopted radial bin ($r_i$, $r_e$ and $r_m$,
  respectively), with the mean radius computed as the average distance
  from the center of all the stars in the bin ($N$, fourth
  column). The last two columns list the second velocity moment and
  its uncertainty in each bin, respectively.}
\label{tab:2ndv_01}
\end{deluxetable*}

\subsection{Systemic rotation}
\label{sec_vrot}
In previous kinematics analysis \citep[e.g.,][]{sollima+19,
  vasiliev+21} no unambiguous signals of rotation have been detected
in the external regions of NGC 6440.  However, thanks to the large
sample of MUSE and SINFONI data presented here, we have the
opportunity to perform the very first exploration of the central
region of the cluster searching for evidence of systemic rotation.

To this purpose, we used the method fully described in previous works
\citep[e.g., see][]{cote+95, lane+09, bellazzini+12, lanzoni+13} and adopted in several
papers \citep{bianchini+13, boberg+17, ferraro+18b, lanzoni+18a,
lanzoni+18b, leanza+22}.  Following this method, the RV sample is split into
two sub-samples by a line passing through the cluster center, varying
the position angle (PA) of the line from $0\arcdeg$ (North direction)
to $180\arcdeg$ (South direction), by steps of $10\arcdeg$ and with PA
$=90\arcdeg$ corresponding to the East direction.  For each value of
PA, the difference between the mean velocity of the two RV sub-samples
($\Delta V_{\rm mean}$) is computed.  In the presence of systemic
rotation, $\Delta V_{\rm mean}$ would show a coherent sinusoidal
variation as a function of PA.  The maximum/minimum absolute value of
this curve provides twice the rotation amplitude ($A_{\rm rot}$) and
the position angle of the rotation axis (PA$_0$). In addition, if
the cluster is rotating, the stellar distribution in a diagram showing
the velocity $V_r$ as a function of the projected distances from the
rotation axis (XR) shows an asymmetry, with two diagonally opposite
quadrants being more populated than the other two.  Moreover, the
sub-samples of stars on each side of the rotation axis are expected to
also show different cumulative $V_r$ distributions.  Three estimators
have been used to quantify the statistical significance of the
detected differences: the probability that the RV distributions of the
two sub-samples are extracted from the same parent family is estimated
by means of a Kolmogorov-Smirnov test, while the statistical
significance of the difference between the two sample means are
evaluated with both the Student's t-test and a Maximum-Likelihood
approach.

Of course, by construction, the method can be used only in the case of
a RV sampling symmetrically distributed in the plane of the sky.  Thus
in order to avoid some heavy under-sampled regions, we were forced to
limit our analysis to the innermost $5\arcsec$ portion (approximately
covering the core radius of the cluster $r_c=6.4\arcsec$,
\citealt{pallanca+21}), where the combination of the MUSE and SINFONI
samples offers a reasonably symmetric coverage of the cluster (see
Figure \ref{fig:mappa}).  We thus performed the analysis over the
entire region ($r<5\arcsec$) and in two radial annuli around the cluster center
($r<3\arcsec$ and $<3\arcsec<r<5\arcsec$).

\begin{figure}[ht!]
\includegraphics[width=20cm, height=12cm]{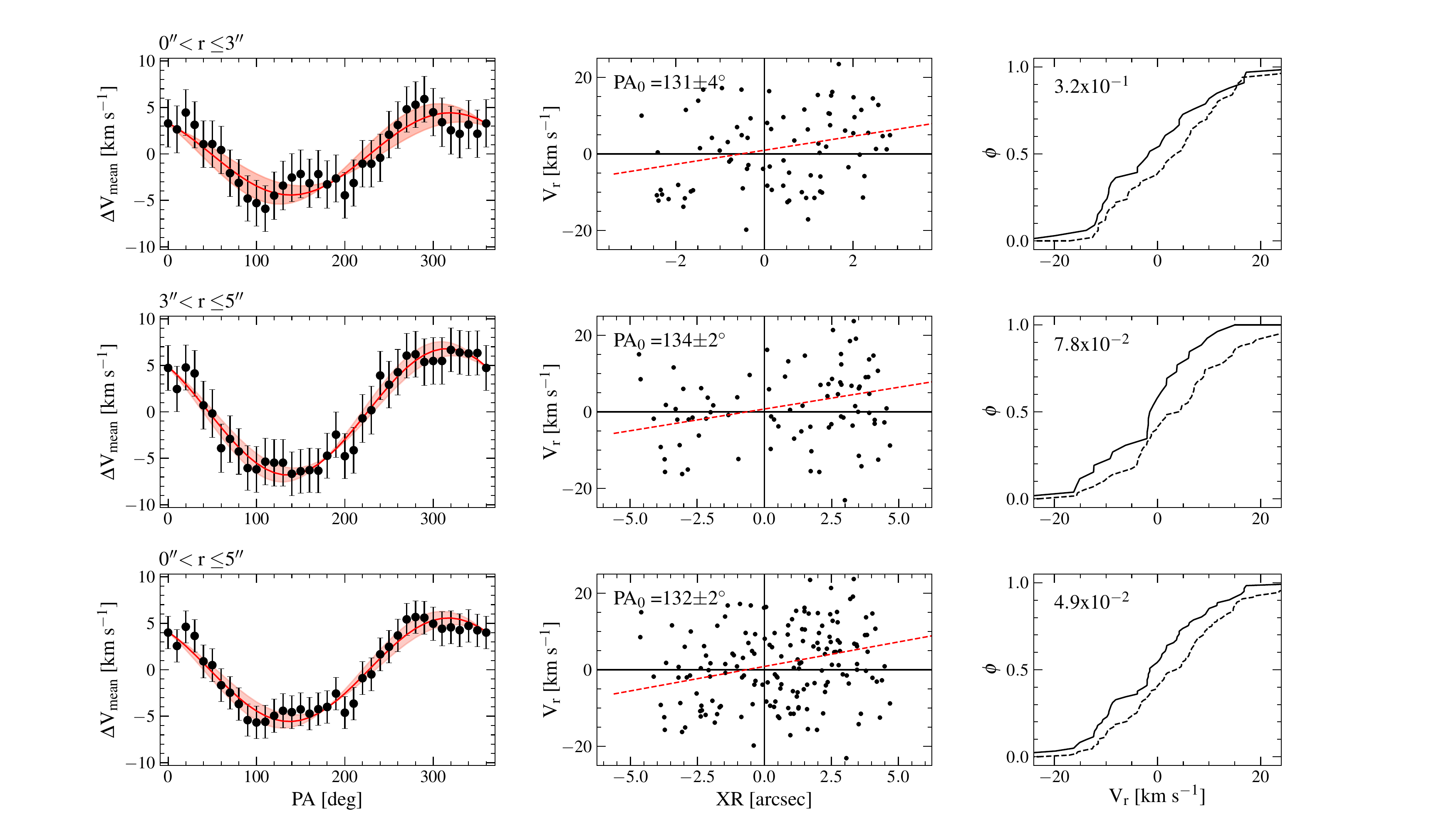}
\centering
\caption{Diagnostic diagrams of the rotation signature detected in
  three concentric radial bins in the core of the cluster at different
  distances from the center (see labels in the top-left corner of each
  row). For each bin, the {\it left panels } show the difference
  between the mean RV on each side of a line passing through the
  center with a given PA, as a function of PA itself. The continuous
  line is the sine function that best fits the observed patterns and
  the red shaded region marks the confidence level at 3$\sigma$. The
  {\it central panels} show the distribution of the velocities $V_r$
  as a function of the projected distances from the rotation axis (XR)
  in arcseconds. The value of PA$_0$ is labeled in each panel.  The
  red dashed lines are the least square fits the data.  The {\it right
    panels} show the cumulative $V_r$ distributions for the stars with
  XR$<0$ (solid line) and for those with XR$>0$ (dotted line). The
  Kolmogorov-Smirnov probability that the two samples are extracted
  from the same parent distribution is also labelled.}
\label{fig:fig_vrot_bin}
\end{figure}

The results are plotted in Figure \ref{fig:fig_vrot_bin} and listed in
Table \ref{tab_vrot_bin}.  The diagnostic plots show the
characteristics of systemic rotation in the considered regions:
a sinusoidal behavior of $\Delta V_{\rm mean}$ as a function of PA
(left-hand panels), asymmetric distributions of $V_r$ as a function of
XR (central panels), and well distinct cumulative $V_r$ distributions
for the two samples on either side of the rotation axis (right-hand
panels). 
Hence, we can reasonably (at $\sim 2\sigma$ statistical
significance) conclude that the core region within $5\arcsec$ 
of NGC 6440 is rotating, with an average position angle of the rotation axis 
PA$_0\sim 132 \pm 2 \arcdeg$, and an amplitude of $\sim 2.8 \pm 0.2$ km s$^{-1}$.
Unfortunately, the non-uniform coverage of the
intermediate region of the cluster does not allow us to assess of the
exact radial extension of the rotation signal. 
Moreover, we have searched for signatures of systemic rotation in outermost part of the cluster by applying the same procedure to the regions covered by KMOS and FLAMES. In this case, no significant evidence of rotation was found, in agreement with previous studies \citep[]{sollima+19, vasiliev+21}.

\begin{deluxetable*}{rrrrccccc}[ht!]
\tablecaption{Rotation signature detected in the core of NGC 6440 in
    three circular annuli around the cluster center.}
\tablewidth{0pt}
\renewcommand{\arraystretch}{1.2}
\tablehead{
\colhead{$r_i$ } & \colhead{ $r_e$ } & \colhead{ $r_m$ } & \colhead{ $N$}  &  \colhead{PA$_0$}  &
\colhead{ $A_{\rm rot}$} & \colhead{ $P_{\rm KS}$}  & \colhead{ $P_{\rm Stud}$} & \colhead{n-$\sigma_{\rm ML}$}
}
\startdata
  0.01  & 3.00  & 2.10 &  85 & $131 \pm 4$ & $2.2 \pm 0.3$ & $3.2\times 10^{-1}$ &  $<90.0$  & 2.2  \\ 
  3.00  & 5.00  & 3.90 &  86 & $134 \pm 2$  & $3.4 \pm 0.3$ & $7.8\times 10^{-2}$ &  $>95.0$  & 3.3  \\ 
  0.01  & 5.00  & 3.00 & 171 & $132 \pm 2$ & $2.8 \pm 0.2$ & $4.9\times 10^{-2}$ &  $>95.0$  & 3.6  \\
\enddata
\tablecomments{The table lists: inner ($r_i$), outer ($r_e$) and mean
  radius ($r_m$) in arcseconds, the number of stars in the bin ($N$),
  the position angle of the rotation axis (PA$_0$) and its 1$\sigma$ 
  error in degree, the rotation
  amplitude (A$_{\rm rot}$) and its error in km s$^{-1}$,
  the Kolmogorov-Smirnov probability that
  the two samples on each side of the rotation axis are drawn from the
  same parent distribution ($P_{\rm KS}$), the t-Student probability
  that the two RV samples have different means ($P_{\rm Stud}$), and
  the significance level (in units of n-$\sigma$) that the two means
  are different following a Maximum-Likelihood approach
  (n-$\sigma_{\rm ML}$).}
\label{tab_vrot_bin}
\end{deluxetable*}

\subsection{Ellipticity}
\label{sec_ellip}
\begin{figure}[ht!]
\includegraphics[width=10.5cm, height=9cm]{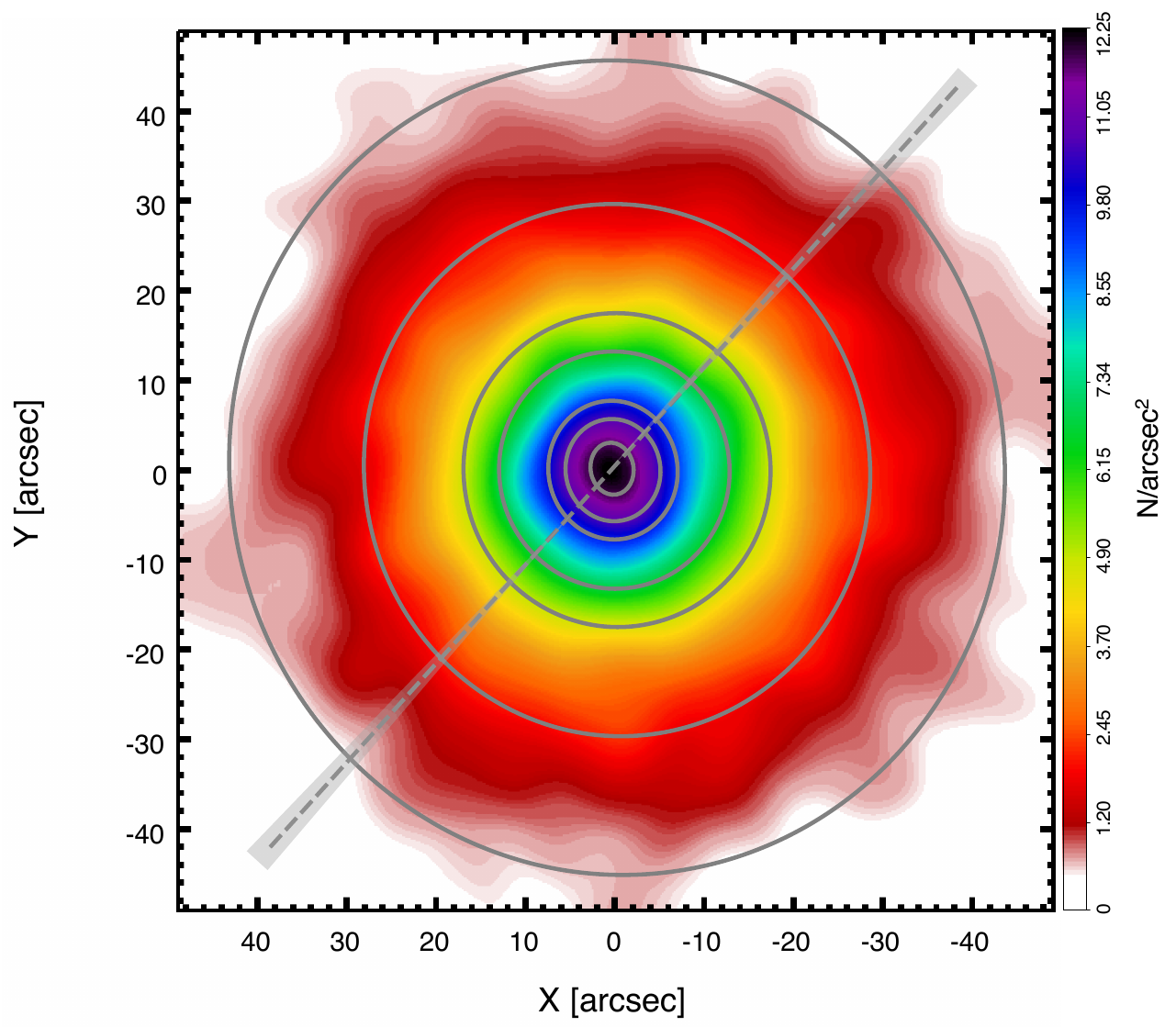}
\centering
\caption{Projected stellar density map of the central region of NGC
  6440. The solid gray lines are the best-fit ellipses to the isodensity curves and the gray dashed line marks the estimated direction of the rotation axis
  (PA$_0 \sim 132 \pm 2\arcdeg$; see Section \ref{sec_vrot}) with the 1$\sigma$ error shown by the shaded area.}
\label{fig:ellipticity}
\end{figure}

From previous analyses, NGC 6440 shows a low global ellipticity
\citep[see][0.01]{harris+96}, as expected for a non-rotating
system. However, because of the rotation signal detected in the core,
we have explored the morphology of the innermost region of the cluster
to check for a possible flattening of the system in the direction
perpendicular to the rotation axis, as expected from theoretical
models \citep[e.g.,][]{Chandrasekhar+69, varri+12}.  We thus
determined the 2D stellar density map of the inner $\sim 50\arcsec
\times 50\arcsec$ area, by using the stars with $I<20.5$ (to avoid
incompleteness effects) in the photometric catalog of
\citet{pallanca+21}. By applying a Gaussian kernel to the stellar
distribution \citep[see][]{dalessandro+15}, we obtained the smoothed
2D surface density map shown in Figure \ref{fig:ellipticity}. The
gray solid lines represent the best-fit ellipses to the isodensity 
contours, and show that
the system is slightly flattened in the center, and it acquirers a
more spherical symmetric for increasing radii.  Indeed, the resulting
ellipticity (defined as $\epsilon = 1 - b/a$ with $a$ and $b$ being
the major and minor axes, respectively) achieves its maximum value
($0.18 \pm 0.02$) at $r \sim 3\arcsec$ and gradually decreases at
larger radii ($\epsilon = 0.04 \pm 0.02$ at $r \sim 45\arcsec$).
Where the ellipticity is maximum, the ellipses major axis has an
orientation of $\sim 15 \arcdeg$ from North to East. Although the
direction of the major axis is not exactly perpendicular to the
rotation axis (PA$_0$ $ \sim 132 \pm 2 \arcdeg$), the presence and the orientation of
the detected flattening are qualitatively consistent with the systemic
rotation signal found in the inner $5\arcsec$ of the system.

\section{Discussion and Conclusions}
\label{sec:discussion}
As part of the ESO-VLT MIKiS Survey \citep{ferraro+18a, ferraro+18b},
here we presented the velocity dispersion profile and the detection of
a core rotation for the bulge GC NGC 6440. Thanks to a combination
of different spectroscopic datasets acquired with appropriate spatial
resolution, we measured the RV of $\sim$1800 individual stars sampling
the entire radial extension, beyond $r_t$ of the cluster.
Using only the sample of member stars properly selected 
with the most reliable RV measures,
we have derived the
velocity dispersion profile of the system from its innermost
($\sim 0.1\arcsec$) to the outer regions of the cluster 
($\sim 250\arcsec$), covering about $5 \times r_h$ from the center.
We also detected a rotation signal
in the innermost $\sim 5\arcsec$ of the cluster, while no conclusions
can be drawn for larger distances because of a non homogeneous
sampling of the more external areas.

The present work complements the detailed photometric study of the
cluster recently performed by our group \citet{pallanca+19,
pallanca+21}, which provided a high-resolution extinction maps in
the direction of the cluster, a new determination of the average
reddening, a new star density profile, and updated values of the
structural parameters, distance modulus and age of the cluster.
Adopting some of these quantities and the resulting value 
of $\sigma_0$ ($12.0 \pm 0.4$ km s $^{-1}$, see Section \ref{sec:2ndv}),
we have verified that
the adopted King model reproduces reasonably well also the velocity
dispersion profile of the cluster (see Figure \ref{fig:vdp}). 

Compared to \citet{Baumgardt+18}, our profile samples a 
much inner region of the cluster, and our estimate of $\sigma_0$
is significantly smaller (12 km s $^{-1}$, in place of
15.8 km s $^{-1}$). This difference could be ascribed 
to an effect of energy equipartition and mass segregation, 
which implies a lower velocity dispersion for RGB stars, 
compared to the average, less massive, cluster members. 
Indeed, based on its age and current half-mass relaxation time 
\citep[$t=13$ Gyr and $t_{rh} = 1$ Gyr, respectively;][see also \citealt{Baumgardt+18}]{pallanca+21},
NGC 6440 has undergone about 13 relaxations so far, and likely is mass-segregated.
Consistently, the value of $\sigma_0$ estimated in this work (which has been obtained mainly from giant stars) is smaller than the mass-weighted central velocity dispersion determined by \citet{Baumgardt+18} from the comparison with N-body simulations.

According to equation (1), in the presence of rotation, the velocity
dispersion is expected to be smaller than the second velocity moment.
However, in the case of NGC 6440 the contribution of the detected
rotation signal is small compared to the measured second velocity
moment, thus we can conclude that the cluster is, with a reasonable
approximation, a pressure-supported system, dominated by non-ordered
motions.  These considerations make the approximation that the cluster
is well represented by single-mass, spherical, isotropic, and
non-rotating \citep{king+66} models well acceptable.  Hence, we can
use the derived value of $\sigma_0$
to estimate the total mass of the system. To this end, we used
equation (3) of \citet{Majewski+03}, deriving $\mu$ as in
\citet{Djorgovski+93} and assuming $\beta = 1/\sigma_0^2$
\citep[see][]{Richstone+86}. We then estimated the total mass
uncertainty by running 1000 Monte Carlo simulations, extracting the
values of $c$, $r_0$ and $\sigma_0$ from a appropriate normal
distributions for each parameter \citep[see][]{leanza+22}.  
The result obtained for NGC 6440 is 
$M = 2.66_{-0.24}^{+0.27} \times 10^5 M_\odot$. 
We emphasize, however, that this value likely underestimates the true total mass of the system, because the adopted central velocity dispersion has been measured from giant stars (see above). This is qualitatively in agreement with the larger mass ($4.42 \pm 0.64 \times 10 ^5$ M$_\odot$) estimated by
\citet{Baumgardt+18} from the mass-weighted value of $\sigma_0$.

Although no evidence of systemic rotation have been detected in
previous analyses \citep[see][]{sollima+19, vasiliev+21}, and although
in the present work we could not derive a rotation curve for reasons
of non-uniform sampling, a not negligible signal of ordered rotation
has been found in the core of NGC 6440 (see Section
\ref{sec_vrot}). 
By assuming that the maximum peak of the rotation is 
$\sim 3.4 \pm 0.3$ km s$^{-1}$ between $3\arcsec$ and $5\arcsec$ from the center,
as found in the present work, 
we can derive the value of $V_{rot}/\sigma_0 = 0.3$, 
similarly to what has been done in other works 
\citep{bianchini+13, fabricius+14, boberg+17, dalessandro+21}.
Nevertheless, since the non-uniform sampling prevents
the exploration of the rotation signal at larger radii,
we cannot exclude that the rotation peak is higher for $r>5\arcsec$.
Despite this, it is interesting to note that such a central rotation is a rare feature in GCs and only two similar cases are known to date in the literature, namely M 15 \citep{vandenBosch+06,usher+21} and NGC 6362 \citep{dalessandro+21}. Interestingly, $N$-body simulations \citep{tiongco+17} show that such central ($<r_h$) velocity signals are expected only in very dynamically evolved systems that lost a significant amount of their mass because of both two-body relaxation effects and interactions with the host galaxy potential. 
However, we stress that this result deserves further investigation and additional analysis at larger radii to see if the rotation signal extends beyond the core region.


\vskip1truecm
S.L. warmly thanks S. Kamann for useful suggestions about PampelMuse.
E.V. acknowledges the Excellence Cluster ORIGINS Funded by the Deutsche Forschungsgemeinschaft (DFG, German Research Foundation) under Germany's Excellence Strategy – EXC-2094 – 390783311.
This work is part of the project {\it Cosmic-Lab} at the Physics and
Astronomy Department "A. Righi" of the Bologna University
(http://www.cosmic-lab.eu/ Cosmic-Lab/Home.html). The research was
funded by the MIUR throughout the PRIN-2017 grant awarded to the
project {\it Light-on-Dark} (PI:Ferraro) through contract PRIN-2017K7REXT.


\newpage

\bibliographystyle{aasjournal}



\end{document}